\documentclass[prd,aps,twocolumn,showpacs,preprintnumbers,amsmath,amssymb]{revtex4}

\usepackage{amsmath,amssymb,graphicx}
\usepackage {amssymb}
\usepackage{multirow}
\usepackage{color}
\newcommand{\nc}{\newcommand}
\nc{\ba}{\begin{eqnarray}} \nc{\ea}{\end{eqnarray}}
\nc{\be}{\begin{equation}} \nc{\ee}{\end{equation}}

\newcommand\s{\sigma}

\newcommand\e{\epsilon}

\newcommand\om{\omega}
\newcommand\ei{\eta_I}
\newcommand\te{\theta}

\nc{\ga}{\gamma} \nc{\x}{{\bf x }} \nc{\kk}{{\bf k }} \nc{\f}{{\bf f
}} \nc{\T}{ \theta (s_i (t)- \s) } \nc{\TT}{ \theta (s_i (t_{ r \, i
} )- \s) } \nc{\br}{   (s_i (t)- \s)  } \nc{\fa}{\phi_1}
\nc{\fb}{\phi_2}

\input{epsf.sty}

\begin{document}

\markright{CYCU-HEP-11-05}

\title{On the Instability of the Lee-Wick Bounce}

\author{ Johanna Karouby$^1$\footnote{karoubyj@mx0.hep.physics.mcgill.ca}, Taotao Qiu$^2$\footnote{xsjqiu@gmail.com} and Robert Brandenberger$^1$\footnote{rhb@mx0.hep.physics.mcgill.ca}}

\affiliation{$^1$Department of Physics, McGill University,
Montr\'eal,
QC, H3A 2T8, Canada\\
$^2$Department of Physics, Chung-Yuan Christian University, Chung-li
320, Taiwan}


\begin{abstract}

It was recently realized \cite{Cai:2008qw} that a model constructed
from a Lee-Wick type scalar field theory yields, at the level of
homogeneous and isotropic background cosmology, a bouncing
cosmology. However, bouncing cosmologies induced by pressure-less
matter are in general unstable to the addition of relativistic
matter (i.e. radiation). Here we study the possibility of obtaining
a bouncing cosmology if we add radiation coupled to the Lee-Wick
scalar field. This coupling in principle would allow the energy
to flow from radiation to matter, thus providing a drain for the
radiation energy. However, we find that it takes an extremely
unlikely fine tuning of the initial phases of the field configurations
for a sufficient amount of radiative energy to flow into matter.
For general initial conditions, the evolution leads to a singularity
rather than a smooth bounce.

\end{abstract}

\maketitle
\section{Introduction}

Both Standard \cite{Hawking:1973uf} and Inflationary
Cosmology \cite{Borde:1993xh} suffer from the initial
singularity problem and hence cannot yield complete
descriptions of the very early universe. If one were able
to construct a non-singular bouncing cosmology, this
problem would obviously disappear. However, in order
to have a chance to obtain such a non-singular cosmology,
one must either go beyond Einstein gravity as a theory of
space-time (see e.g. \cite{Mukhanov:1991zn} for an early
construction), or else one must make use of matter which
violates the ``null energy condition" (see \cite{Novello}
for a review of both types of approaches).

Interest in non-singular bouncing cosmologies has
increased with the realization that they
can lead to alternatives to inflationary
cosmology as a theory for the origin of structure in the
universe. A specific scenario which can arise
at the level of homogeneous and isotropic
cosmology is the ``matter bounce" paradigm which is
based on the realization  \cite{Wands, Finelli}
that vacuum fluctuations which exit the Hubble radius during
a matter-dominated contracting phase evolve into
a scale-invariant spectrum of curvature perturbations on
super-Hubble scales before the bounce. The key
point is that the curvature fluctuation variable $\zeta$ grows
on super-Hubble scales in a contracting phase, whereas
it is constant on these large scales in an expanding phase.
Since long wavelength modes exit the Hubble radius
earlier than short wavelength ones, they grow for a longer
period of time. This provides a mechanism for reddening the
initial vacuum spectrum. It turns out that a matter dominated
contracting phase provides the specific boost in the
power of long wavelength modes which is required in
order to transform a vacuum spectrum into a scale-invariant
one. Studies in the case of various non-singular bounce models
\cite{models} have shown that on wavelengths long compared to the
duration of the bounce phase, the spectrum of fluctuations is
virtually unchanged during the bounce. Thus, a scale-invariant
spectrum of curvature fluctuations survives on super-Hubble
scales at late times.

Provided that the bounce can occur at energy scales much below
the Planck scale, non-singular cosmologies solve a key conceptual
problem from which inflationary cosmology suffers, namely the
``Trans-Planckian" problem for fluctuations \cite{RHBrev0, Martin:2000xs}:
If the period of inflationary expansion of space lasts for more than
$70 H^{-1}$, where $H$ is the Hubble expansion rate during inflation
(in order to solve the key cosmological mysteries it was designed to
explain, inflation has to last at least $50 H^{-1}$), then the physical
wavelengths of even the largest-scale
fluctuation modes we see today will be even smaller than the Planck
length at the beginning of inflation and thus in the ``zone of ignorance"
where the physics on which inflation and the theory of cosmological
perturbations are based, namely Einstein gravity coupled to semiclassical
field theory matter, will break down. In contrast, in a non-singular
bouncing cosmology the wavelength of modes which are currently
probed by cosmological observations is never much smaller than
$1 {\rm mm}$ (the physical wavelength of the mode which corresponds
to our current Hubble radius evaluated when the temperature of the
universe was $10^{16} {\rm GeV}$) and hence many orders of
magnitude larger than the Planck length. Thus, the fluctuations
never enter the ``trans-Planckian zone of ignorance" of sub-Planck-length
wavelengths.

Possibly the simplest realization of the matter bounce scenario is
the ``quintom bounce" model \cite{Cai:2007qw} and is obtained by
considering the matter sector to contain two scalar fields, one of them
(the ``ghost field") having the ``wrong" sign of the kinetic action.
The potential of the ghost scalar field also has the opposite sign to
that of regular scalar fields such that in the absence of interactions,
the ghost field has a classically stable minimum. As has been
noticed in \cite{Cai:2008qw}, such a quintom bounce model
also arises from the scalar field sector of the ``Lee-Wick" (LW)
Lagrangian \cite{LW} which contains higher derivatives terms.

The quintom and Lee-Wick bouncing cosmologies are obtained in
the following way \cite{Cai:2007qw, Cai:2008qw}: We begin in
the contracting phase with both the regular and the ghost scalar
field oscillating homogeneously in space about their respective
vacua. We assume that the energy density is dominated by the
regular matter field, and that hence the total energy density is
positive. Once the amplitude of the regular scalar field exceeds the
Planck scale, the field oscillations will freeze out and a slow-climb
phase will begin during which the energy density of the field
only grows slowly (this is the time reverse of the slow-roll
phase in scalar field-driven inflation). However, the ghost field
continues to oscillate and its energy density (which is negative)
continues to grow in absolute value. Hence, the total energy
density drops to zero, at which point the bounce occurs, as has
been studied both analytically and numerically in the above-mentioned
works. Note that the energy density in this bounce model
scales as matter until the regular scalar field freezes out.

A major problem of bouncing cosmologies realized with matter which
scales as $a^{-3}$ as a function of the scale factor $a(t)$ is the
potential instability of the homogeneous and isotropic background
against the effects of radiation (which scales as $a^{-4}$ and
anisotropic stress which scales as $a^{-6}$ \footnote{One of the
major advantages of the Ekpyrotic bouncing scenario \cite{Ekp} is
that the contracting phase is stable against such effects.}. If we
simply add a non-interacting radiation component to the two scalar
field system, then unless the initial energy density in radiation is
tuned to be extremely small, then the radiation component will
become dominant long before the bounce can arise, and will prevent
the energy density in the ghost field from ever being able to become
important, resulting in a Big Crunch singularity. Similarly, unless
the initial energy density in anisotropic stress is very small, it
will come to dominate the energy density of the universe long before
the bounce is expected. The anisotropies will destabilize the
homogeneous background cosmology, and will prevent a bounce. Note
that at the quantum level, there is an additional severe problem for
bounce models obtained with matter fields with ghost-like kinetic
terms, namely the quantum instability of the vacuum (see e.g.
\cite{Jeon}).

In this paper we will focus on the radiation instability problem.
For the purpose of this discussion we will simply assume that
anisotropic stress is absent. In a recent paper, two of us studied
the possibility that a bounce could arise if radiation is
supplemented with Lee-Wick radiation \cite{Karouby:2010wt}. However,
we showed that this hope is not realized: the addition of Lee-Wick
radiation does not prevent the Big Crunch singularity from
occurring. In the presence of radiation, the only hope to obtain a
bounce is to introduce a coupling between radiation and ghost scalar
field matter which could effectively drain energy density from the
radiation field and prevent the energy density of radiation from
becoming dominant.  Here we study this possibility. However, at
least for the specific Lagrangian which we consider, we find that a
bounce only emerges for highly fine-tuned phases of the fields and
their velocities in the initial conditions.

The paper is organized as follows: In Section II, we introduce the
model we study, namely the scalar field sector of Lee-Wick theory
coupled to radiation, and write down the general
equations of motion.  In Section III
we set up the equations of motion linearized about the bounce
background, treating the entire radiation field as an inhomogeneous
fluctuation. In particular, we study the different terms which
contribute to the energy-momentum tensor and identify those which
could assist in obtaining a non-singular bounce. In Section IV we
study the solutions of the perturbed equations of motion, and in
Section V we analyze the evolution of the different terms in the
energy-momentum tensor, identifying the conditions which would be
required in order to obtain a non-singular bounce. We have also
evolved the general equations of motion for the two inhomogeneous
scalar field configurations and the classical inhomogeneous
radiation field in the homogeneous background cosmology. Section VI
summarizes some of the numerical results. Both the analytical and
numerical results confirm that we need unnatural fine-tuning of the
initial conditions in order to obtain a non-singular bounce. In the
final section we offer some conclusions and discussion.

\section{The Model}

The Lee-Wick scalar field model coupled to electromagnetic
radiation is given by the following Lagrangian:
\ba\label{leewick} {\cal
L}&=&-\frac{1}{2}\partial_{\mu}\phi\partial^{\mu}\phi+\frac{1}{2M^2}(\partial^2
\phi)^2-\frac{1}{2}m^2\phi^2-V(\phi) \nonumber \\
&&-\frac{1}{ 4} F_{\mu \nu}
F^{\mu\nu}-f(\phi,\partial^2\phi,F_{\mu\nu}F^{\mu\nu})~, \ea
where $m$ is the mass of the scalar field $\phi$, and $V(\phi)$ is its
potential. Here we adopt the convention that
\be
ds^2 \, = \, -dt^2 + a^2(t)(dx^2 + dy^2 + dz^2) \, ,
\ee
where $a(t)$ is the scale factor of the universe. Since it is a
higher derivative Lagrangian in $\phi$,  the scalar field
sector contains an extra degree of freedom
with the ``wrong" sign kinetic term and with a
mass set by the scale $M$. We choose  $m\leq M\leq m_{Pl}$,
where $m_{Pl}$ is the Planck mass, since we want the regular
scalar field to dominate at low energies, but at the same time
we do not want to worry about quantum gravity effects.
The second line of the Lagrangian
(\ref{leewick}) contains the kinetic term of the radiation as well
as the coupling term, where we assumed both for the sake of
generality and because of foresight that the
radiation field couples not only to the scalar field $\phi$ itself,
but also to the higher derivative term. The electromagnetic tensor,
$F_{\mu\nu}$, is related to the radiation field $A_\mu$ through the
usual definition
\be F_{\mu\nu} \, \equiv \, \nabla_\mu A_\nu-\nabla_\nu A_\mu \, ,
\ee 
where $\nabla_\mu$ is the covariant derivative.

It is convenient to extract the extra degree of freedom as a
separate scalar field. To do this, we use the field redefinitions
\ba
\phi \, &\equiv& \, \phi_1-\phi_2 \, , \nonumber \\
\phi_2 \, &\equiv& \, \partial^2 \phi/M^2 \, .
\ea
The Lagrangian (\ref{leewick}) then takes on
a simpler form:
\ba\label{leewick2} {\cal L} \, &=& \,
-\frac{1}{2}\partial_{\mu}\phi_1\partial^{\mu}\phi_1+\frac{1}{2}\partial_{\mu}\phi_2\partial^{\mu}\phi_2-\frac{1}{2}m^2
\fa^2+\frac{1}{2}M^2 \fb^2\, \nonumber\\
&& \, -\frac{1}{ 4} F_{\mu \nu} F^{\mu\nu}-f(\phi_1,\phi_2,F_{\mu\nu}F^{\mu\nu})~,
\ea
where we have chosen the potential to be zero. In this new form, the Lagrangian
describes two massive scalar fields with one of them (i.e., $\phi_2$) behaving
like a``ghost", and both of them coupled to the radiation field.

The coupling term $f(\phi_1,\phi_2,F_{\mu\nu}F^{\mu\nu})$ should in
principle be arbitrary, however, in this paper we will take a
specific form for convenience. The form will be:
\be \label{F1}
f(\phi_1,\phi_2,F_{\mu\nu}F^{\mu\nu}) \, = \, \frac{1}{4}
(c\fa^2+d\fb^2)F_{\mu \nu} F^{\mu \nu}~,
\ee
where $c$ and $d$ are coupling constants which have mass dimension
$-2$. The interaction terms are non-renormalizable. To make sure
that such terms could be thought of as arising from an effective
field theory which is consistent at the bounce, we must make sure
that the coefficients are chosen such that the contribution of
the interaction term to the Lagrangian density is smaller than
that of the other terms. This must be true even at energy densities
at which the bounce occurs in the pure scalar field model. It is
easy to see that this condition will be satisfied if the coefficients
$c$ and $d$ are both of the order $m_{pl}^{-2}$.

It is the purpose of this paper to study the effects which these
coupling terms have on the dynamics of the system. We know that in
the absence of coupling, i.e. when $c=d=0$, a bounce will only occur
if the initial radiation energy density is tuned to a very small
value compared to the scalar field energy density. This is because
the positive definite energy density of radiation will scale as
$a^{-4}$ which is faster than that of the scalar fields, in
particular the ghost scalar field. Generically, it will dominate the
energy of the universe after some amount of contraction, it will
prevent the ghost scalar field energy density from catching up and
will thus prevent a bounce, leading to a Big Crunch singularity
instead. With non-vanishing values of $c$ and $d$, however, the
scalars are in principle able to drain energy from the radiation.

From the Lagrangian (\ref{leewick2}), one can obtain the stress-energy
tensor $T_{\mu\nu}$ by varying the action with respect to the metric
$g^{\mu\nu}$. In the hydrodynamical limit, we can take
$T_{\mu\nu}$ to be of the form of
$diag\{\rho,a^2(t)p_1,a^2(t)p_2,a^2(t)p_3\}$
where $\rho$ and $p$ are energy density and pressure, respectively. For now,
we consider the general form which is:
\begin{widetext}
\ba\label{tmunu} T_{\mu\nu}&=&g_{\mu\nu}{\cal
L}+\partial_\mu\phi_1\partial_\nu\phi_1-\partial_\mu\phi_2\partial_\nu\phi_2+(1-c\phi_1^2-d\phi_2^2)F_{\mu\lambda}F_\nu^\lambda~,\nonumber\\
&=&g_{\mu\nu}\bigl[\frac{\dot\phi_1^2}{2}-\frac{1}{2a^2}\partial_i\phi_1\partial_i\phi_1-\frac{1}{2}m^2\phi_1^2
-\frac{\dot\phi_2^2}{2}+\frac{1}{2a^2}\partial_i\phi_2\partial_i\phi_2+\frac{1}{2}M^2\phi_2^2-\frac{1}{4}(1-c\phi_1^2-d\phi_2^2)F^2\bigr]~\nonumber\\
&&+\partial_\mu\phi_1\partial_\nu\phi_1-\partial_\mu\phi_2\partial_\nu\phi_2+(1-c\phi_1^2-d\phi_2^2)F_{\mu\lambda}F_\nu^\lambda~.
\ea
\end{widetext}
Since we will be studying the contribution of plane wave perturbations of
the scalar fields and we will treat radiation as a superposition of waves,
we kept the space-derivative terms.

By varying the Lagrangian with respect to the matter fields
$\phi_1$, $\phi_2$ and $A_\mu$, we also get the equations of motion
for all three fields: 
\ba
\label{phi1eom}&&\Box\phi_1-(m^2-\frac{c}{2}F^2)\phi_1=0~,\\
\label{phi2eom}&&\Box\phi_2-(M^2+\frac{d}{2}F^2)\phi_2=0~,\\
\label{amueom}&&(1-c\phi_1^2-d\phi_2^2)(\partial_\nu
F^{\mu\nu}+3HF^{\mu0})\nonumber\\
&&-2(c\phi_1\partial_\nu\phi_1+d\phi_2\partial_\nu\phi_2)F^{\mu\nu}=0~
\ea
which will be analyzed in detail in the rest of the paper.

\section{Dynamics}

Since the equations of motion are nonlinear, we cannot work in
Fourier space, and use plane wave solutions. However, we are
interested in how initially small amounts of radiation build
up and possibly transfer their energy to scalar field
fluctuations. We treat radiation as a superposition of fluctuations.
Therefore it makes sense to linearize our equations about the
homogeneous scalar field background. Thus, we make the
following ansatz for the scalar fields:
\ba\label{ansatzphi}
\phi_1(t,z)&=&\phi_{1}^{(0)}(t)+\e \phi_{1}^{(1)}(t,z)+\e^2 \phi_{1}^{(2)}(t)  \\
\phi_2(t,z)&=&\phi_{2}^{(0)}(t)+\e \phi_{2}^{(1)}(t,z)+\e^2
\phi_{2}^{(2)}(t)~,
\ea
where the expansion parameter $\e$ is taken to be much smaller than
$1$ \footnote{The expansion parameter $\e$ should be viewed as
parameterizing the initial ratio of radiation energy to background
scalar field energy. Thus, the leading contribution of the radiation
field is first order in $\e$. Via the coupling terms in the
Lagrangian with coefficients $c$ and $d$, the linear radiation field
induces linear scalar field inhomogeneities $\phi_{1}^{(1)}$ and
$\phi_{2}^{(1)}$. These corrections will contain a further
suppression factor since $c$ and $d$ are small coefficients.
Similarly, the same coupling terms in the Lagrangian will lead to a
perturbation $\delta \gamma$ of the rescaled radiation field
$\gamma$ which is of linear order in $\e$ but suppressed by factors
of $c$ and $d$.}. The first term on the right hand side of each
line, i.e. $\phi_{1,2}^{(0)}(t)$ correspond to the background
fields, the terms $\phi_{1,2}^{(1)}(t, x)$ are the fluctuations, and
the second order terms $\phi_{1,2}^{(2)}(t)$ describe the
back-reaction of the fluctuations on the background and can be
computed from the leading second order corrections (averaged over
space) of the equations of motion \footnote{By taking the scalar
product of the second order equations with a fixed plane wave
(instead of averaging over space) one could also compute the
back-reaction of the fluctuations on the inhomogeneous modes.}.

To simplify the analysis, we describe radiation in terms of plane waves
in a fixed direction (which we take to be the $z$ direction).
Without loss of generality we can restrict
attention to one polarization mode which we take to be the electric
field in the $x$ direction and the magnetic field in the $y$
direction. In this case, the only non-zero components of the field
strength tensor are $F^{01}$ and $F^{13}$. Using the temporal gauge
where $A_0 =0$, we find that only the first component of the gauge
field is non-zero. For a single wavelength fluctuation
we can make the ansatz
\be\label{ansatzgauge1}
A_1(k,t) \, = \, f(t)cos(kz) \equiv\gamma(k,t)~,
\ee
or, equivalently,
\be\label{ansatzgauge2}
A^1(k,t) \, = \, a(t)^{-2} \gamma(k,t)~.
\ee
Since in the linearized equations of motion the Fourier modes are independent,
we can consider $\phi_{1}^{(1)}$ and $\phi_{2}^{(1)}$
also to be plane waves propagating in $z$ direction, so they depend only
on $z$ and $t$.

With Eqs. (\ref{ansatzphi}-\ref{ansatzgauge2}) in hand, we can write
down the energy densities of the various fields at each order in
perturbation theory.

\subsection{The stress-energy tensor}

First of all, we insert the above perturbative ansatz for the fields into the
stress-energy tensor of the system. From the general expression
(\ref{tmunu}) for $T_{\mu\nu}$ we get:
\begin{widetext}\ba\label{tmunu2}
T_{\mu\nu}&=&g_{\mu\nu}\bigl[\frac{\dot\phi_1^2}{2}-\frac{1}{2a^2}\partial_z\phi_1\partial_z\phi_1-\frac{1}{2}m^2\phi_1^2
-\frac{\dot\phi_2^2}{2}+\frac{1}{2a^2}\partial_z\phi_2\partial_z\phi_2+\frac{1}{2}M^2\phi_2^2-\frac{1}{4}(1-c\phi_1^2-d\phi_2^2)F^2\bigr]~\nonumber\\
&&+\partial_\mu\phi_1\partial_\nu\phi_1-\partial_\mu\phi_2\partial_\nu\phi_2+(1-c\phi_1^2-d\phi_2^2)F_{\mu\lambda}F_\nu^\lambda~.
\ea
\end{widetext}

The $00$ component of Eq. (\ref{tmunu2}) denotes the energy density
of the system:
\ba\label{rho}
\rho&=&\frac{1}{2}(\dot\phi_1^2+\frac{k^2}{a^2}\phi_{1}^2+m^2\phi_1^2)
-\frac{1}{2}(\dot\phi_2^2+\frac{k^2}{a^2}\phi_{2}^2+M^2\phi_2^2)\nonumber\\
&&+(1-c\phi_1^2-d\phi_2^2)(\frac{F^2}{4}+F_{0\lambda}F_0^{\lambda})~,
\ea
so at each level in perturbation theory we have:
\ba
\label{rho0}\rho^{(0)}&=&\frac{1}{2}(\dot{\phi_1^{(0)}}^2+m^2{\phi_1^{(0)}}^2)
-\frac{1}{2}(\dot{\phi_2^{(0)}}^2+M^2{\phi_2^{(0)}}^2)~,\\
\label{rho1}\rho^{(1)}&=&(\dot\phi_1^{(0)}\dot\phi_1^{(1)}+m^2\phi_1^{(0)}\phi_1^{(1)})
-(\dot\phi_2^{(0)}\dot\phi_2^{(1)}\nonumber\\
&&+M^2\phi_2^{(0)}\phi_2^{(1)})~,\\
\label{rho2}\rho^{(2)}&=&\frac{1}{2}(\dot{\phi_1^{(1)}}^2+\dot\phi_1^{(0)}\dot\phi_1^{(2)}
+\frac{k^2}{a^2}{\phi_{1}^{(1)}}^2+m^2{\phi_1^{(1)}}^2\nonumber\\
&&+m^2\phi_1^{(0)}\phi_1^{(2)})-\frac{1}{2}(\dot{\phi_2^{(1)}}^2+\dot\phi_2^{(0)}\dot\phi_2^{(2)}
+\frac{k^2}{a^2}{\phi_{2}^{(1)}}^2\nonumber\\
&&+M^2{\phi_2^{(1)}}^2+M^2\phi_2^{(0)}\phi_2^{(2)})\nonumber\\
&&+(1-c{\phi_1^{(0)}}^2-d{\phi_2^{(0)}}^2)(\frac{F^2}{4}+F_{0\lambda}F_0^{\lambda})~,\nonumber\\
&=&\frac{1}{2}(\dot{\phi_1^{(1)}}^2+\dot\phi_1^{(0)}\dot\phi_1^{(2)}
+\frac{k^2}{a^2}{\phi_{1}^{(1)}}^2+m^2{\phi_1^{(1)}}^2\nonumber\\
&&+m^2\phi_1^{(0)}\phi_1^{(2)})-\frac{1}{2}(\dot{\phi_2^{(1)}}^2+\dot\phi_2^{(0)}\dot\phi_2^{(2)}
+\frac{k^2}{a^2}{\phi_{2}^{(1)}}^2\nonumber\\
&&+M^2{\phi_2^{(1)}}^2+M^2\phi_2^{(0)}\phi_2^{(2)})\nonumber\\
&&+(1-c{\phi_1^{(0)}}^2-d{\phi_2^{(0)}}^2)(\frac{k^2}{2a^4}\gamma^2+\frac{\dot\gamma^2}{2a^2})~.
\ea

We can similarly obtain the pressure of the system from the $ii$
components of Eq. (\ref{tmunu2}). Note that due to the anisotropy in
$T_{\mu\nu}$ caused by the gauge field as well as by the anisotropic
fluctuations of the scalar fields, the pressures in the three
directions are no longer identical. The pressure in each direction
can be written as:
\ba\label{pres}
p_i&=&\frac{1}{2}(\dot\phi_1^2-\frac{k^2}{a^2}\phi_{1}^2-m^2\phi_1^2)
-\frac{1}{2}(\dot\phi_2^2-\frac{k^2}{a^2}\phi_{2}^2-M^2\phi_2^2)\nonumber\\
&&-(1-c\phi_1^2-d\phi_2^2)(\frac{F^2}{4}-\frac{F_{i\lambda}F_i^{\lambda}}{a^2})+\frac{\partial_i\phi_1\partial_i\phi_1}{a^2}\nonumber\\
&&-\frac{\partial_i\phi_2\partial_i\phi_2}{a^2}~,
\ea
with no summation over the index $i$. From this formula, we can see that at
both zero-th and first order, the pressure is isotropic:
\ba
p_i^{(0)}&=&\frac{1}{2}(\dot{\phi_1^{(0)}}^2-m^2{\phi_1^{(0)}}^2)
-\frac{1}{2}(\dot{\phi_2^{(0)}}^2-M^2{\phi_2^{(0)}}^2)~,\\
p_i^{(1)}&=&(\dot\phi_1^{(0)}\dot\phi_1^{(1)}-m^2\phi_1^{(0)}\phi_1^{(1)})
-(\dot\phi_2^{(0)}\dot\phi_2^{(1)}\nonumber\\
&&-M^2\phi_2^{(0)}\phi_2^{(1)})~,
\ea
while the second order pressure for each direction gives
\ba
p_i^{(2)}&=&\frac{1}{2}(\dot{\phi_1^{(1)}}^2+\dot\phi_1^{(0)}\dot\phi_1^{(2)}
-\frac{k^2}{a^2}{\phi_{1}^{(1)}}^2+m^2{\phi_1^{(1)}}^2\nonumber\\
&&+m^2\phi_1^{(0)}\phi_1^{(2)})-\frac{1}{2}(\dot{\phi_2^{(1)}}^2+\dot\phi_2^{(0)}\dot\phi_2^{(2)}
-\frac{k^2}{a^2}{\phi_{2}^{(1)}}^2\nonumber\\
&&+M^2{\phi_2^{(1)}}^2+M^2\phi_2^{(0)}\phi_2^{(2)})
+\frac{\partial_i\phi_1\partial_i\phi_1}{a^2}-\frac{\partial_i\phi_2\partial_i\phi_2}{a^2}\nonumber\\
&&-(1-c{\phi_1^{(0)}}^2-d{\phi_2^{(0)}}^2)(\frac{F^2}{4}-\frac{F_{i\lambda}F_i^{\lambda}}{a^2})~,
\ea
where $i=1,2,3$.

We can thus obtain every component of $p_i^{(2)}$:
\ba
p_1^{(2)}&=&\frac{1}{2}(\dot{\phi_1^{(1)}}^2+\dot\phi_1^{(0)}\dot\phi_1^{(2)}
-\frac{k^2}{a^2}{\phi_{1}^{(1)}}^2+m^2{\phi_1^{(1)}}^2\nonumber\\
&&+m^2\phi_1^{(0)}\phi_1^{(2)})-\frac{1}{2}(\dot{\phi_2^{(1)}}^2+\dot\phi_2^{(0)}\dot\phi_2^{(2)}
-\frac{k^2}{a^2}{\phi_{2}^{(1)}}^2\nonumber\\
&&+M^2{\phi_2^{(1)}}^2+M^2\phi_2^{(0)}\phi_2^{(2)})\nonumber\\
&&+(1-c{\phi_1^{(0)}}^2-d{\phi_2^{(0)}}^2)(\frac{k^2}{2a^4}\gamma^2-\frac{\dot\gamma^2}{2a^2})~,\\
p_2^{(2)}&=&\frac{1}{2}(\dot{\phi_1^{(1)}}^2+\dot\phi_1^{(0)}\dot\phi_1^{(2)}
-\frac{k^2}{a^2}{\phi_{1}^{(1)}}^2+m^2{\phi_1^{(1)}}^2\nonumber\\
&&+m^2\phi_1^{(0)}\phi_1^{(2)})-\frac{1}{2}(\dot{\phi_2^{(1)}}^2+\dot\phi_2^{(0)}\dot\phi_2^{(2)}
-\frac{k^2}{a^2}{\phi_{2}^{(1)}}^2\nonumber\\
&&+M^2{\phi_2^{(1)}}^2+M^2\phi_2^{(0)}\phi_2^{(2)})\nonumber\\
&&-(1-c{\phi_1^{(0)}}^2-d{\phi_2^{(0)}}^2)(\frac{k^2}{2a^4}\gamma^2-\frac{\dot\gamma^2}{2a^2})~,\\
p_3^{(2)}&=&\frac{1}{2}(\dot{\phi_1^{(1)}}^2+\dot\phi_1^{(0)}\dot\phi_1^{(2)}
-\frac{k^2}{a^2}{\phi_{1}^{(1)}}^2+m^2{\phi_1^{(1)}}^2\nonumber\\
&&+m^2\phi_1^{(0)}\phi_1^{(2)})-\frac{1}{2}(\dot{\phi_2^{(1)}}^2+\dot\phi_2^{(0)}\dot\phi_2^{(2)}
-\frac{k^2}{a^2}{\phi_{2}^{(1)}}^2\nonumber\\
&&+M^2{\phi_2^{(1)}}^2+M^2\phi_2^{(0)}\phi_2^{(2)})\nonumber\\
&&+(1-c{\phi_1^{(0)}}^2-d{\phi_2^{(0)}}^2)(\frac{k^2}{2a^4}\gamma^2+\frac{\dot\gamma^2}{2a^2})~,
\ea
and the average is:
\ba
p_{eff}^{(2)}&=&\frac{p_1^{(2)}+p_2^{(2)}+p_3^{(2)}}{3}~\\
&=&\frac{1}{2}(\dot{\phi_1^{(1)}}^2+\dot\phi_1^{(0)}\dot\phi_1^{(2)}
-\frac{k^2}{a^2}{\phi_{1}^{(1)}}^2+m^2{\phi_1^{(1)}}^2\nonumber\\
&&+m^2\phi_1^{(0)}\phi_1^{(2)})-\frac{1}{2}(\dot{\phi_2^{(1)}}^2+\dot\phi_2^{(0)}\dot\phi_2^{(2)}
-\frac{k^2}{a^2}{\phi_{2}^{(1)}}^2\nonumber\\
&&+M^2{\phi_2^{(1)}}^2+M^2\phi_2^{(0)}\phi_2^{(2)})\nonumber\\
&&+(1-c{\phi_1^{(0)}}^2-d{\phi_2^{(0)}}^2)(\frac{k^2}{6a^4}\gamma^2+\frac{\dot\gamma^2}{6a^2})~.
\ea

From the above, we can also see that in order to analyze the behavior of
the energy density up to second order, we need to know the evolution
of scalar fields up to second order as well as that of the
gauge field up to first order, while the behavior of the gauge field
to second order is not required.

It will be useful in the following to separate the contributions to
the energy density and pressure in a different way, namely, \\
i) the contribution from the background homogeneous part of the scalar
fields,
\be
\rho_{\phi}^{h} =
\rho_{\phi}^{(0)}\,\ ~~~ p_{\phi}^{h}= p_{\phi}^{(0)}~,
\ee
ii) that of the scalar field perturbations (in slight abuse of
notation we call this the ``inhomogeneous'' term),
\be
\rho_{\phi}^{inh} = \e
\rho_{\phi}^{(1)}+\e^2 \rho_{\phi}^{(2)}\, ~~~
p_{\phi}^{inh} = \e p_{\phi}^{(1)}+\e^2 p_{\phi}^{(2)}~,
\ee
iii) the contribution of the gauge field,
\ba \rho_g &=&\frac{1}{2a^2}(\dot
\ga^2+\frac{k^2}{a^2}\ga^2)~,\\
p_g &=&\frac{1}{12a^2}(\dot \ga^2+\frac{k^2}{a^2}\ga^2)~,
\ea
and iv) the contribution of the coupling term,
\ba
\rho_c &= &-\frac{(c {\phi_1^{(0)}}^2+d{\phi_2^{(0)}}^2)}{8 a^2}(\dot \ga^2+\frac{k^2}{a^2}\ga^2)=-\Phi \rho_g~,\\
p_c&=&-\frac{(c{\phi_1^{(0)}}^2+d {\phi_2^{(0)}}^2)} {24 a^2}(\dot
\ga^2+\frac{k^2}{a^2}\ga^2)=-\Phi p_g~,
\ea
where in the last equation we define $\Phi$ to be the quadratic
combination of the two fields:
\be
\Phi = (c{\fa^{(0)}}^2 + d {\fb^{(0)}}^2)/2 \, .
\ee

{F}rom the above we can deduce the equation of state parameter for
each part:
\ba
w_{\phi}^{h}&=&\frac{\dot{{\phi_1^{(0)}}}^2-m^2{\phi_1^{(0)}}^2-\dot{{\phi_2^{(0)}}}^2
+M^2{\phi_2^{(0)}}^2}{\dot{{\phi_1^{(0)}}}^2+m^2{\phi_1^{(0)}}^2-\dot{{\phi_2^{(0)}}}^2
-M^2{\phi_2^{(0)}}^2}~,\\ w_{\phi}^{inh}&=&\e\left(
\frac{p^{(1)}}{\rho^{(1)}}-\frac{p^{(0)}\rho^{(1)}}{{\rho^{(0)}}^2}
\right)\nonumber\\
&&+\e^2\left(\frac{p^{(2)}}{\rho^{(0)}}-\frac{p^{(0)}\rho^{(2)}}{{\rho^{(0)}}^2}-\frac{p^{(1)}\rho^{(1)}}{{\rho^{(0)}}^2}
\right)~,\\
 w_g&=&w_c=\frac{1}{3}~.
\ea

From the equations above, we see that for positive values of the
constants $c$ and $d$, the coupling of the scalar
field with the gauge field  will give rise to a contribution $\rho_c$
to the energy density which has the same equation of state but opposite
sign to that of the gauge field. Therefore the coupling can help drain
energy from the gauge field. It is because of this mechanism that we
might hope to achieve a cosmological bounce in the presence of
radiation. A first indication on whether a bounce might occur
can be obtained by considering the scaling of each contribution
to the energy density as a function of the scale factor $a(t)$.
To find these scalings, we need the time dependence of the
linear and quadratic contributions to each field. Therefore,
we need to solve the matter field equations of motion. In the following
subsection, we present the equations for the fields at each order,
while the solutions and detailed analysis will be performed in the
next sections.

\subsection{Equations of Motion}

Keeping in mind the ans\"atze for $A_\mu$, $\phi_1$ and $\phi_2$,
their equations of motion at each order can be obtained from
(\ref{phi1eom}), (\ref{phi2eom}) and (\ref{amueom}): \\
a) At zero-th order: 
\ba\label{eqn0}
\left\{
\begin{array}{c} \ddot\phi_1^{(0)} + 3H
\dot\phi_1^{(0)} + m^2\phi_1^{(0)} = 0~,\\
\\
\ddot\phi_2^{(0)} + 3H\dot\phi_2^{(0)} + M^2\phi_2^{(0)} = 0~. \\
\end{array}\right.
\ea
Note that there is no equation at this order for $A_\mu$ because it
is of first order in $\e$. \\
b) At first order:
\ba\label{eqn1}
\left\{ \begin{array}{c} \ddot\phi_1^{(1)} + 3H
\dot\phi_1^{(1)} + (\frac{k^2}{a^2}+m^2)\phi_1^{(1)} = 0~,\\
\\
\ddot\phi_2^{(1)} + 3H\dot\phi_2^{(1)} + (\frac{k^2}{a^2}+M^2)\phi_2^{(1)} = 0~,\\
\\
(1-c{\phi_1^{(0)}}^2 - d{\phi_2^{(0)}}^2)(\partial_\nu
F^{\mu\nu}+3HF^{\mu0})
\\
- 2(c\phi_1^{(0)}\partial_\nu\phi_1^{(0)} + d\phi_2^{(0)}\partial_\nu\phi_2^{(0)})F^{\mu\nu}=0~,\\
\end{array}\right.
\ea
Making use of Eqs. (\ref{ansatzgauge1}) and (\ref{ansatzgauge2}),
the equation for the gauge field can also be rewritten as:
\ba\label{max}
&&(1-c{\phi_1^{(0)}}^2 - d{\phi_2^{(0)}}^2)(\ddot\gamma + H\dot\gamma + \frac{k^2}{a^2}\gamma)\nonumber\\
&&- 2(c\phi_1^{(0)}\dot\phi_1^{(0)} + d\phi_2^{(0)}\dot\phi_2^{(0)})\dot\gamma = 0~.
\ea
c) At second order:
\ba \label{eqn2}
\left\{ \begin{array}{c} \ddot\phi_1^{(2)} + m^2\phi_1^{(2)}
-\frac{c}{2}<F_{\mu\nu}F^{\mu\nu}>\phi_1^{(0)} = 0~,\\
\\
\ddot\phi_2^{(2)} + M^2\phi_2^{(2)} +
\frac{d}{2}<F_{\mu\nu}F^{\mu\nu}>\phi_2^{(0)} = 0~.\\
\end{array}\right.
\ea
Here, pointed parentheses indicate spatial averaging (since we are only
focusing on the zero mode of the second order field fluctuations). We
also neglected the effect of Hubble friction since it does not give an
important contribution for the second order fluctuations.

\section{The general solution}

In this section we will solve the equations of motion (\ref{eqn0}),
(\ref{eqn1}) and (\ref{eqn2}) to see if and how a bounce will
happen.

It is usually useful to perform the analysis in the conformal frame
where the conformal time $\eta\equiv\int a^{-1}(t)dt$ is used rather
than the cosmic time. Additionally, to extract the dependence on
the scale factor, it is convenient to use the following two variables:
\be
u_1(\eta) \equiv a(\eta)\phi_1(\eta),~~~u_2(\eta) \equiv
a(\eta)\phi_2(\eta)~.
\ee
Hereafter, we will use $u^{(i)}_j(i=0,1,2,~j=1,2)$ to denote the
$i$-th order perturbation of the $j$-th scalar field. Moreover,
for simplicity but without loosing generality, we can parameterize
the scale factor $a(t)$ as
\be \label{para}
a(\eta) = a_0t^p = a_0|\eta|^{\frac{p}{1-p}}~,
\ee
with
\be
p = \frac{2}{3(1+w)}~,
\ee
where $a_0$ and $w$ are the initial value of
the scale factor and the equation of state of the universe,
respectively. This is a self-consistent assumption when $w$ is
nearly a constant. The evolution of $w$ in our case will be shown
numerically in Section VI.

\subsection{Solutions for $\phi_1^{(0)}$ and $\phi_2^{(0)}$}
\label{subsecA}

Using the parametrization (\ref{para}), the equations of motion at
zeroth order of the two scalar fields become:
\ba\label{eqn0new}
\left\{
\begin{array}{c} {u_1^{(0)}}''+(a_0^2 m^2 \eta^\frac{2p}{1-p}-\frac{p(2p-1)}{(1-p)^2\eta^2})u_1^{(0)}=0~,\\
\\
{u_2^{(0)}}''+(a_0^2 M^2 \eta^\frac{2p}{1-p}-\frac{p(2p-1)}{(1-p)^2\eta^2})u_2^{(0)}=0~,\\
\end{array}\right.
\ea
where a prime denotes the derivative with respect to conformal time $\eta$.
Their solutions are:
\ba
u_1^{(0)} &\sim& \sqrt{|\eta|}H_{\pm\frac{1-3p}{2}}\bigl((1-p)am|\eta|\bigr)~,\\
u_2^{(0)} &\sim& \sqrt{|\eta|}H_{\pm\frac{1-3p}{2}}\bigl((1-p)aM|\eta|\bigr)~,
\ea
where $H_{\pm\frac{1-3p}{2}}$ represents the $(\pm\frac{1-3p}{2})$-th
order Hankel function. Far away or close to the bounce, i.e. for
$a|\eta|\gg m^{-1},M^{-1}$ and $a|\eta|\ll m^{-1},M^{-1}$, respectively,
the approximate solutions are: \\
1)Oscillations for large values of the scale factor $a|\eta|\gg m^{-1},M^{-1}$:
\ba \label{sol0sub1}
u_1^{(0)}&\sim&|\eta|^\frac{p}{2(p-1)}\sqrt{\frac{2}{(1-p)\pi
a_0m}}\cos\bigl((1-p)a_0m|\eta|^\frac{1}{1-p}\nonumber\\
&&+\te_1^{(0)}\bigr)~,\\
\label{sol0sub2}
u_2^{(0)}&\sim&|\eta|^\frac{p}{2(p-1)}\sqrt{\frac{2}{(1-p)\pi
a_0M}}\cos\bigl((1-p)a_0M|\eta|^\frac{1}{1-p}\nonumber\\
&&+\te_2^{(0)}\bigr)~,
\ea
(where $\te_1$ and $\te_2$ are phases set by the initial conditions). In
terms of the non-rescaled fields $\phi_i$ one obtains damped (or anti-damped)
oscillations (depending on whether we are in an expanding or a contracting
period)
\ba \label{sol0phisub1}
\phi_1^{(0)}&\sim&|\eta|^\frac{3p}{2(p-1)}\sqrt{\frac{2}{(1-p)\pi
a^\frac{3}{2}_0m}}\cos\bigl((1-p)a_0m|\eta|^\frac{1}{1-p}\nonumber\\
&&+\te_1^{(0)}\bigr)~,\\
\label{sol0phisub2}
\phi_2^{(0)}&\sim&|\eta|^\frac{3p}{2(p-1)}\sqrt{\frac{2}{(1-p)\pi
a^\frac{3}{2}_0M}}\cos\bigl((1-p)a_0M|\eta|^\frac{1}{1-p}\nonumber\\
&&+\te_2^{(0)}\bigr)~.
\ea
2)``Frozen'' evolution for small values of the scale factor $a|\eta|\ll m^{-1},M^{-1}$:
\ba \label{sol0sup1}
u_1^{(0)}&\sim&\frac{\bigl((1-p)a_0m\bigr)^{\frac{1-3p}{2}}|\eta|^\frac{1-2p}{1-p}}{\Gamma(\frac{3(1-p)}{2})}\nonumber\\
&&+\frac{\bigl((1-p)a_0m\bigr)^{-\frac{1-3p}{2}}|\eta|^\frac{p}{1-p}}{\Gamma(\frac{1+3p}{2})}~,\\
\label{sol0sup2} u_2^{(0)}&\sim&\frac{\bigl((1-p)a_0M\bigr)^{\frac{1-3p}{2}}|\eta|^\frac{1-2p}{1-p}}{\Gamma(\frac{3(1-p)}{2})}\nonumber\\
&&+\frac{\bigl((1-p)a_0M\bigr)^{-\frac{1-3p}{2}}|\eta|^\frac{p}{1-p}}{\Gamma(\frac{1+3p}{2})}~,
\ea
from which it follows that the non-rescaled fields $\phi_i$ evolve as
\ba \label{sol0phisup1}
\phi_1^{(0)}&\sim&\frac{\bigl((1-p)a_0m\bigr)^{\frac{1-3p}{2}}|\eta|^\frac{1-3p}{1-p}}{a_0\Gamma(\frac{3(1-p)}{2})} \nonumber \\
& & +\frac{\bigl((1-p)a_0m\bigr)^{\frac{3p-1}{2}}}{a_0\Gamma(\frac{1+3p}{2})}~,\\
\label{sol0phisup2}
\phi_2^{(0)}&\sim&\frac{\bigl((1-p)a_0M\bigr)^{\frac{1-3p}{2}}|\eta|^\frac{1-3p}{1-p}}{a_0\Gamma(\frac{3(1-p)}{2})} \nonumber \\
&& +\frac{\bigl((1-p)a_0M\bigr)^{\frac{3p-1}{2}}}{a_0\Gamma(\frac{1+3p}{2})}~,
\ea
from which we can see that the last term of $\phi_i^{(0)}$ is a
constant mode while the first term is a varying one. Depending on
the value of $p$ (or equivalently $w$) the varying mode could be
growing (for $p>1/3$ or $-1<w<1$), in which case it becomes dominant,
or decaying (for $p<1/3$ or $w>1/w<-1$), in which case it becomes
subdominant. We can usually neglect the decaying part of the fields.

\subsection{Solutions for $\phi_1^{(1)}$ and $\phi_2^{(1)}$}

Following the steps performed in the last subsection, we can also
get the solutions for the first order components of the scalar fields.
Using the equations (\ref{eqn1}) for the first order perturbations
we obtain the following equations of motion for $u_1^{(1)}$ and $u_2^{(1)}$:
\ba \label{eqn1new} \left\{
\begin{array}{c} {u_1^{(1)}}''+(k^2+a_0^2 m^2 \eta^\frac{2p}{1-p}-\frac{p(2p-1)}{(1-p)^2\eta^2})u_1^{(1)}=0~,\\
\\
{u_2^{(1)}}''+(k^2+a_0^2 M^2 \eta^\frac{2p}{1-p}-\frac{p(2p-1)}{(1-p)^2\eta^2})u_2^{(1)}=0~.\\
\end{array}\right.
\ea

Depending on the value of $k$, we obtain different
approximation solutions. For wavenumbers large compared both
to the Hubble radius and to the mass term, we obtain oscillatory
solutions with fixed amplitude.

Considering now modes which are still sub-Hubble (i.e. $k|\eta| >
1$) but for which the mass term dominates over the contribution of
the field tension (i.e. the term involving $k$), we can neglect both
the $k^2$ term and the term involving
$\frac{p(2p-1)}{(1-p)^2\eta^2}$. The simplified equation for these
modes is: 
\ba \left\{\label{u1in}
\begin{array}{c} {u_1^{(1)}}'' + a_0^2 m^2 \eta^\frac{2p}{1-p} u_1^{(1)} = 0~,\\
\\
{u_2^{(1)}}'' + a_0^2 M^2 \eta^\frac{2p}{1-p} u_2^{(1)} = 0~,\\
\end{array}\right.
\ea
whose solutions are:
\ba
u_1^{(1)}&\sim&\sqrt{|\eta|}H_{\frac{1-p}{2}}\bigl((1-p)a_0m|\eta|\bigr)\nonumber\\
&\sim&|\eta|^{\frac{p}{2(p-1)}}\sqrt{\frac{2}{(1-p)\pi
ma_0}}\cos\bigl((1-p)a_0m|\eta|\nonumber\\
&&+\theta^{(1)}_1\bigr)~,\\
u_2^{(1)}&\sim&\sqrt{|\eta|}H_{\frac{1-p}{2}}\bigl((1-p)a_0M|\eta|\bigr)\nonumber\\
&\sim&|\eta|^{\frac{p}{2(p-1)}}\sqrt{\frac{2}{(1-p)\pi
Ma_0}}\cos\bigl((1-p)a_0M|\eta|\nonumber\\
&&+\theta^{(1)}_2\bigr)~.\ea

For modes outside the Hubble radius ($k\eta\ll 1$), we have:
\ba \left\{
\begin{array}{c} {u_1^{(1)}}'' + (a_0^2 m^2 \eta^\frac{2p}{1-p}-\frac{p(2p-1)}{(1-p)^2\eta^2})u_1^{(1)} = 0~,\\
\\
{u_2^{(1)}}'' + (a_0^2 M^2 \eta^\frac{2p}{1-p}-\frac{p(2p-1)}{(1-p)^2\eta^2})u_2^{(1)} = 0~,\\
\end{array}\right.
\ea
which have the same form as Eq. (\ref{eqn0new}) so their solution
will be the same as given in Eqs. (\ref{sol0sup1}) and (\ref{sol0sup2}).

We have thus seen that the first order solutions for the scalar
fields scale the same way with $|\eta|$ as the zeroth order
solution. This is because in the small $|\eta|$ region where the
$a''(t)/a(t)$ term dominates over the other ones, the equations for
first order and zero-th order modes are almost the same. Thus,
unless the energy density in the $u_i^{(1)}$ modes dominates at the
initial time, it will never dominate over the background
contribution from the $u_i^{(0)}$ terms. Thus we can conclude that
the first order fluctuations of scalar fields will not prevent the
bounce.

\subsection{Solution for the gauge field $\gamma$}

In this section, we will analyze the gauge field $\gamma$ which is
also considered to be of first order. The equation (\ref{max}) can
directly be transformed to conformal frame as:
\be \label{gammaeq}
\gamma'' + k^2\gamma - \frac{2(c\phi_1^{(0)}{\phi_1^{(0)}}'+d\phi_2^{(0)}{\phi_2^{(0)}}')}{1-c{\phi_1^{(0)}}^2-d{\phi_2^{(0)}}^2}\gamma' = 0~.
\ee
Since the coefficients $c$ and $d$ are small, we can take the last term
to be a source term. In a first order Born approximation, we can
write the total solution as
\be
\gamma \simeq \gamma_0 + \delta\gamma \, ,
\ee
where $\gamma_0$ is the solution for the homogeneous equation obtained by
setting $c = d = 0$, while $\delta\gamma$ is the leading correction term
obtained by inserting $\gamma_0$ into the source term (the last term in
(\ref{gammaeq})).

The zero-th order (homogeneous) equation is easily solved and gives
\be \label{gamma}
\gamma_0 \sim \cos(k|\eta| + \te_\gamma)~.
\ee

For the first order equation, it is convenient to define
\be
P(\eta) \equiv -\frac{2(c\phi_1^{(0)}{\phi_1^{(0)}}'+d\phi_2^{(0)}{\phi_2^{(0)}}')}{1-c{\phi_1^{(0)}}^2-d{\phi_2^{(0)}}^2} \, ,
\ee
so that the equation becomes
\be
\delta \gamma'' + k^2 \delta \gamma + P(\eta)\gamma_0' = 0~,
\ee
where we neglected the small term $P(\eta)\delta\gamma'$. Inserting the
solution of $\gamma_0$ (\ref{gamma}), we get the following
equation for $\delta\gamma$:
\be \label{maxnew}
\delta\gamma'' + k^2\delta\gamma = -P(\eta)\gamma_0'~.
\ee

We are interested in the scaling of $\delta \gamma$ as a function of
time. For this purpose, we need to work out the scaling in time of
the source term in (\ref{maxnew}). Since the solutions for $\phi_1^{(0)}$
and $\phi_2^{(0)}$ scale differently in time in the two time intervals
discussed in Subsection (\ref{subsecA}), it is necessary to
analyze these two intervals separately.

For times obeying $a|\eta| \gg m^{-1},M^{-1}$, then by differentiating
(\ref{sol0phisub1}) and (\ref{sol0phisub2}) with respect to $\eta$ we have:
\ba
{\phi_1^{(0)}}' &\sim& |\eta|^\frac{p+2}{2(p-1)}\sqrt{\frac{2}{(1-p)\pi
a^\frac{3}{2}_0m}}\\
&&\times\bigl[-\frac{3p}{p-1}\cos\bigl((1-p)a_0m|\eta|^\frac{1}{1-p}+\te_1^{(0)}\bigr)\nonumber\\
&&+a_0m|\eta|^\frac{1}{1-p}\sin\bigl((1-p)a_0m|\eta|^\frac{1}{1-p}+\te_1^{(0)}\bigr)\bigr]~, \nonumber \\
{\phi_2^{(0)}}'&\sim&|\eta|^\frac{p+2}{2(p-1)}\sqrt{\frac{2}{(1-p)\pi
a^\frac{3}{2}_0M}}\\
&&\times\bigl[-\frac{3p}{p-1}\cos\bigl((1-p)a_0M|\eta|^\frac{1}{1-p}+\te_1^{(0)}\bigr)\nonumber\\
&&+a_0M|\eta|^\frac{1}{1-p}\sin\bigl((1-p)a_0M|\eta|^\frac{1}{1-p}+\te_1^{(0)}\bigr)\bigr]~. \nonumber
\ea
Note that $|\eta|^\frac{1}{1-p}\sim t$ is a decaying
mode in the contracting phase and thus the last terms inside the
square brackets in the above formulae can be neglected compared to
the first ones. Since
\ba
P(\eta) &=& -\frac{2(c\phi_1^{(0)}{\phi_1^{(0)}}'+d\phi_2^{(0)}{\phi_2^{(0)}}')}{1-c{\phi_1^{(0)}}^2-d{\phi_2^{(0)}}^2} \nonumber \\
&\approx& -2(c\phi_1^{(0)}{\phi_1^{(0)}}'+d\phi_2^{(0)}{\phi_2^{(0)}}') \, ,
\ea
then combining all these results we get:
\be
\delta\gamma \sim C_1|\eta|^\frac{1-4p}{1-p}~.
\ee

For $a|\eta|\ll m^{-1},M^{-1}$ , then differentiating (\ref{sol0phisup1}) and (\ref{sol0phisup2})
with respect to $\eta$ we obtain
\ba
{\phi_1^{(0)}}' &\sim& \frac{1-3p}{1-p}\frac{\bigl((1-p)a_0m\bigr)^{\frac{1-3p}{2}}|\eta|^\frac{-2p}{1-p}}{a_0\Gamma(\frac{3(1-p)}{2})}~,\\
{\phi_2^{(0)}}' &\sim& \frac{1-3p}{1-p}\frac{\bigl((1-p)a_0M\bigr)^{\frac{1-3p}{2}}|\eta|^\frac{-2p}{1-p}}{a_0\Gamma(\frac{3(1-p)}{2})}~,
\ea
when $p>1/3$ and
\be
{\phi_1^{(0)}}'\sim{\phi_2^{(0)}}'\approx 0
\ee
when $p<1/3$. Then we can solve Equation (\ref{maxnew}) to get:
\ba
\delta\gamma &\sim& C_2|\eta|^\frac{3-7p}{1-p}~,~~~p>\frac{1}{3} \nonumber\\
\delta\gamma &\sim& C_3\cos(k|\eta|+\theta_{\delta\gamma})~.~~~p<\frac{1}{3}
\ea
In the above expressions for $\delta\gamma$, $C_1$, $C_2$ and $C_3$ are
complicated prefactors in front of the $\eta$-dependent terms.

In summary, we see that the interactions give only a subleading
correction $\delta\gamma$ to $\gamma_0$.

\subsection{Solutions for $\phi_1^{(2)}$ and $\phi_2^{(2)}$}

Finally, let us consider the homogeneous component of the second
order fluctuations of the scalars, namely, $\phi_1^{(2)}$ and
$\phi_2^{(2)}$. If we only consider the energy density up to
second order, these second order field perturbations give a
contribution through their coupling to the background fields.
In the following we find the solutions of (\ref{eqn2}) and study
the effects of the induced terms in the stress-energy tensor on a
possible bounce.

Given the solution for the gauge field $\gamma$ obtained in the last
subsection, it is easy to rewrite Eqs. (\ref{eqn2}) as:
\ba\label{eqn2new} \left\{ \begin{array}{c}
{u_1^{(2)}}'' + a^2m^2u_1^{(2)} = c\frac{(k^2\gamma^2-\gamma'^2)}{a^2}u_1^{(0)}~,\\
\\
{u_2^{(2)}}'' + a^2M^2u_2^{(2)} = -d\frac{(k^2\gamma^2-\gamma'^2)}{a^2}u_2^{(0)}~,\\
\end{array}\right.
\ea
where we made use of the fact that
\be
F_{\mu\nu}F^{\mu\nu} = 2(k^2\gamma^2-a^2\dot\gamma^2)/a^4 \, .
\ee

Equation (\ref{eqn2new}) has the same form as the zero-th order
equation but with a small source term generated by the interaction
with the gauge field. This equation can be solved using the Born
approximation (details are given in the Appendix). The general
solution is the sum of the general solution of the homogeneous
solution plus the solution including the source which has vanishing
initial data. The inhomogeneous term is suppressed by the coupling
constants $c$ and $d$ compared to the homogeneous solution, but, as
shown in the Appendix, it scales as a high power of $\eta^{-1}$. Via
the coupling to the background scalar fields, the above second order
terms enter into the expression for the energy density to second
order. The signs of the corresponding terms in the energy density are
indefinite in the sense that they depend on the phases of the initial
field configurations. Since it is these terms that dominate the
energy density near the bounce, we find that whether a bounce occurs
or not depends sensitively on the phases in the initial conditions,
and that in fact in the case of many plane wave modes initially
excited, a bounce requires very special phase correlations.

\section{Evolution of the components of the energy density}

In the previous section we have solved all of the field equations
up to second order in the amplitude of the fluctuations. We have found
the scaling in time of each field at each order. Now we are ready
to look at how all of the terms in the expression for the
energy density $\rho^{(0)}$, $\rho^{(1)}$ and $\rho^{(2)}$ at
various orders in perturbation theory
(namely, Eqs. (\ref{rho0})-(\ref{rho2})) scale in time. This
analysis is straightforward but very important if we are to determine
whether a bounce is possible, since in four space-time dimensional
classical Einstein Gravity with flat spatial sections a bounce can
only happen when the negative terms in  the energy density catch up
to the positive contributions \cite{Cai:2007qw}.

In the following we give a table of how each term contained in
$\rho$ scales with time as the background cosmology bounce point
(the bounce which is achieved in the absence of radiation and scalar
field inhomogeneities) is approached. We will identify the terms
which dominate in this limit. This will give us a good indication
under which conditions a bounce can occur. The tables are structured
as follows: the first line ``Terms'', indicates which term we are
considering, the next set of lines ``Behavior'' gives the scaling in
time in the various limits and in the two relevant ranges of the
parameter $p$ which indicates the equation of state, and the last
line gives the sign with which the term contributes to the energy
density. Note that we focus on the growing mode solution to each
field (which is constant for small $\eta$ in the case $p < 1/3$). We
give separate tables for terms of zero-th, first and second order in
$\e$.

a) For terms contained in $\rho^{(0)}$:\\
\begin{widetext}
\begin{tabular}{|c|c|c|c|c|}\hline
Terms & $\dot{\phi_1^{(0)}}^2$ & $m^2{\phi_1^{(0)}}^2$ & $-\dot{\phi_2^{(0)}}^2$ & $-M^2{\phi_2^{(0)}}^2$ \\
\hline Behavior & \begin{tabular}{c}
$a^{-3-\frac{2}{p}}\big(a|\eta|\gg m^{-1}\big)$\\ \hline
$a^{-6}$\big(\begin{tabular}{c} $a|\eta|\ll m^{-1}$\\$p>\frac{1}{3}$
\end{tabular}\big)\\ \hline 0 \big(\begin{tabular}{c} $a|\eta|\ll
m^{-1}$\\$p<\frac{1}{3}$
\end{tabular}\big)
\end{tabular} & \begin{tabular}{c}
$a^{-3}\big(a|\eta|\gg m^{-1}\big)$\\ \hline
$a^{-6+\frac{2}{p}}$\big(\begin{tabular}{c} $a|\eta|\ll
m^{-1}$\\$p>\frac{1}{3}$
\end{tabular}\big)\\ \hline $a^0$ \big(\begin{tabular}{c} $a|\eta|\ll
m^{-1}$\\$p<\frac{1}{3}$
\end{tabular}\big)
\end{tabular} & \begin{tabular}{c}
$a^{-3-\frac{2}{p}}\big(a|\eta|\gg m^{-1}\big)$\\ \hline
$a^{-6}$\big(\begin{tabular}{c} $a|\eta|\ll m^{-1}$\\$p>\frac{1}{3}$
\end{tabular}\big)\\ \hline 0 \big(\begin{tabular}{c} $a|\eta|\ll
m^{-1}$\\$p<\frac{1}{3}$
\end{tabular}\big)
\end{tabular} & \begin{tabular}{c}
$a^{-3}\big(a|\eta|\gg M^{-1}\big)$\\ \hline
$a^{-6+\frac{2}{p}}$\big(\begin{tabular}{c} $a|\eta|\ll
M^{-1}$\\$p>\frac{1}{3}$
\end{tabular}\big)\\ \hline $a^0$ \big(\begin{tabular}{c} $a|\eta|\ll
M^{-1}$\\$p<\frac{1}{3}$
\end{tabular}\big)
\end{tabular}
\\
\hline Sign & \begin{tabular}{c} Positive\\Definite \end{tabular} &
\begin{tabular}{c} Positive\\Definite \end{tabular} &
\begin{tabular}{c} Negative\\Definite \end{tabular} &
\begin{tabular}{c} Negative\\Definite \end{tabular}
\\
\hline
\end{tabular}
\end{widetext}

b) For terms contained in $\rho^{(1)}$:\\
\begin{widetext}
\begin{tabular}{|c|c|c|c|c|}\hline
Terms & $\dot\phi_1^{(0)}\dot\phi_1^{(1)}$ & $\phi_1^{(0)}\phi_1^{(1)}$ & $-\dot\phi_2^{(0)}\dot\phi_2^{(1)}$ \\
\hline Behavior & \begin{tabular}{c}
$a^{-3-\frac{2}{p}}\big(|\eta|\gg Max\{k^{-1},(am)^{-1}\}\big)$\\
\hline $a^{-\frac{9}{2}-\frac{1}{p}}$\big(\begin{tabular}{c}
$|\eta|\in[k^{-1},(am)^{-1}]$\\$p>\frac{1}{3}$ \end{tabular}\big)\\
\hline $a^{-6}$\big(\begin{tabular}{c} $|\eta|\ll
Min\{k^{-1},(am)^{-1}\}$ \\ $p>\frac{1}{3}$ \end{tabular}\big)\\
\hline 0 \big(\begin{tabular}{c} $|\eta|\in[k^{-1},(am)^{-1}]$ \\
$p<\frac{1}{3}$ \end{tabular}\big) \\ \hline 0
\big(\begin{tabular}{c} $|\eta|\ll
Min\{k^{-1},(am)^{-1}\}$\\$p<\frac{1}{3}$
\end{tabular}\big)
\end{tabular} & \begin{tabular}{c}
$a^{-3}\big(|\eta|\gg Max\{k^{-1},(am)^{-1}\}\big)$\\
\hline $a^{-\frac{9}{2}+\frac{1}{p}}$\big(\begin{tabular}{c}
$|\eta|\in[k^{-1},(am)^{-1}]$\\$p>\frac{1}{3}$ \end{tabular}\big)\\
\hline $a^{-6+\frac{2}{p}}$\big(\begin{tabular}{c} $|\eta|\ll
Min\{k^{-1},(am)^{-1}\}$ \\ $p>\frac{1}{3}$ \end{tabular}\big)\\
\hline $a^{-\frac{3}{2}}$ \big(\begin{tabular}{c} $|\eta|\in[k^{-1},(am)^{-1}]$ \\
$p<\frac{1}{3}$ \end{tabular}\big) \\ \hline $a^{0}$
\big(\begin{tabular}{c} $|\eta|\ll
Min\{k^{-1},(am)^{-1}\}$\\$p<\frac{1}{3}$
\end{tabular}\big)
\end{tabular} & \begin{tabular}{c}
$a^{-3-\frac{2}{p}}\big(|\eta|\gg Max\{k^{-1},(aM)^{-1}\}\big)$\\
\hline $a^{-\frac{9}{2}-\frac{1}{p}}$\big(\begin{tabular}{c}
$|\eta|\in[k^{-1},(aM)^{-1}]$\\$p>\frac{1}{3}$ \end{tabular}\big)\\
\hline $a^{-6}$\big(\begin{tabular}{c} $|\eta|\ll
Min\{k^{-1},(aM)^{-1}\}$ \\ $p>\frac{1}{3}$ \end{tabular}\big)\\
\hline 0 \big(\begin{tabular}{c} $|\eta|\in[k^{-1},(aM)^{-1}]$ \\
$p<\frac{1}{3}$ \end{tabular}\big) \\ \hline 0
\big(\begin{tabular}{c} $|\eta|\ll
Min\{k^{-1},(aM)^{-1}\}$\\$p<\frac{1}{3}$
\end{tabular}\big)
\end{tabular}
\\
\hline Sign & Indefinite & Indefinite & Indefinite
\\
\hline
\end{tabular}
\begin{tabular}{|c|c|}\hline
Terms & $-\phi_2^{(0)}\phi_2^{(1)}$ \\ \hline Behavior &
\begin{tabular}{c}
$a^{-3}\big(|\eta|\gg Max\{k^{-1},(aM)^{-1}\}\big)$\\
\hline $a^{-\frac{9}{2}+\frac{1}{p}}$\big(\begin{tabular}{c}
$|\eta|\in[k^{-1},(aM)^{-1}]$\\$p>\frac{1}{3}$ \end{tabular}\big)\\
\hline $a^{-6+\frac{2}{p}}$\big(\begin{tabular}{c} $|\eta|\ll
Min\{k^{-1},(aM)^{-1}\}$ \\ $p>\frac{1}{3}$ \end{tabular}\big)\\
\hline $a^{-\frac{3}{2}}$ \big(\begin{tabular}{c} $|\eta|\in[k^{-1},(aM)^{-1}]$ \\
$p<\frac{1}{3}$ \end{tabular}\big) \\ \hline $a^0$
\big(\begin{tabular}{c} $|\eta|\ll
Min\{k^{-1},(aM)^{-1}\}$\\$p<\frac{1}{3}$
\end{tabular}\big)
\end{tabular}
\\
\hline Sign & Indefinite \\
\hline
\end{tabular}
\end{widetext}
These terms, however, all vanish if the energy density is defined
by spatial averaging.

c) For terms contained in $\rho^{(2)}$:\\
\begin{widetext}
\begin{tabular}{|c|c|c|c|c|c|}\hline
Terms & $\dot{\phi_1^{(1)}}^2$ & $\dot\phi_1^{(0)}\dot\phi_1^{(2)}$
& $\frac{k^2}{a^2}{\phi_{1}^{(1)}}^2$ & $m^2{\phi_1^{(1)}}^2$ &
$m^2\phi_1^{(0)}\phi_1^{(2)}$
\\
\hline Behavior & \begin{tabular}{c}
$a^{-3-\frac{2}{p}}\big(k|\eta|\gg 1\big)$\\ \hline
$a^{-6}$\big(\begin{tabular}{c} $k|\eta|\ll 1$\\$p>\frac{1}{3}$
\end{tabular}\big)\\ \hline 0 \big(\begin{tabular}{c} $k|\eta|\ll
1$\\$p<\frac{1}{3}$
\end{tabular}\big)
\end{tabular} & \begin{tabular}{c}
$a^{-7-\frac{1}{p}}\big(a|\eta|\gg m^{-1}\big)$\\ \hline
$a^{-\frac{17}{2}}$\big(\begin{tabular}{c} $a|\eta|\ll
m^{-1}$\\$p>\frac{1}{3}$
\end{tabular}\big)\\ \hline 0 \big(\begin{tabular}{c} $a|\eta|\ll
m^{-1}$\\$p<\frac{1}{3}$
\end{tabular}\big)
\end{tabular} & \begin{tabular}{c}
$a^{-5}\big(k|\eta|\gg 1\big)$\\ \hline
$a^{-8+\frac{2}{p}}$\big(\begin{tabular}{c} $k|\eta|\ll
1$\\$p>\frac{1}{3}$
\end{tabular}\big)\\ \hline $a^{-2}$ \big(\begin{tabular}{c} $k|\eta|\ll
1$\\$p<\frac{1}{3}$
\end{tabular}\big)
\end{tabular} & \begin{tabular}{c}
$a^{-3}\big(k|\eta|\gg 1\big)$\\ \hline
$a^{-6+\frac{2}{p}}$\big(\begin{tabular}{c} $k|\eta|\ll
1$\\$p>\frac{1}{3}$
\end{tabular}\big)\\ \hline $a^0$ \big(\begin{tabular}{c} $k|\eta|\ll
1$\\$p<\frac{1}{3}$
\end{tabular}\big)
\end{tabular} & \begin{tabular}{c}
$a^{-7+\frac{1}{p}}\big(a|\eta|\gg m^{-1}\big)$\\ \hline
$a^{-\frac{17}{2}+\frac{2}{p}}$\big(\begin{tabular}{c} $a|\eta|\ll
m^{-1}$\\$p>\frac{1}{3}$
\end{tabular}\big)\\ \hline $a^{-\frac{11}{2}+\frac{1}{p}}$ \big(\begin{tabular}{c} $a|\eta|\ll
m^{-1}$\\$p<\frac{1}{3}$
\end{tabular}\big)
\end{tabular}
\\
\hline Sign & \begin{tabular}{c} Positive\\Definite \end{tabular} &
Indefinite & \begin{tabular}{c} Positive\\Definite \end{tabular} &
\begin{tabular}{c} Positive\\Definite \end{tabular} & Indefinite
\\
\hline
\end{tabular}\\
\begin{tabular}{|c|c|c|c|c|c|}\hline
Terms & $-\dot{\phi_2^{(1)}}^2$ &
$-\dot\phi_2^{(0)}\dot\phi_2^{(2)}$ &
$\frac{k^2}{a^2}{\phi_{2}^{(1)}}^2$ & $-M^2{\phi_2^{(1)}}^2$ &
$-M^2\phi_2^{(0)}\phi_2^{(2)}$
\\
\hline Behavior & \begin{tabular}{c}
$a^{-3-\frac{2}{p}}\big(k|\eta|\gg 1\big)$\\ \hline
$a^{-6}$\big(\begin{tabular}{c} $k|\eta|\ll 1$\\$p>\frac{1}{3}$
\end{tabular}\big)\\ \hline 0 \big(\begin{tabular}{c} $k|\eta|\ll
1$\\$p<\frac{1}{3}$
\end{tabular}\big)
\end{tabular} & \begin{tabular}{c}
$a^{-7-\frac{1}{p}}\big(a|\eta|\gg M^{-1}\big)$\\ \hline
$a^{-\frac{17}{2}}$\big(\begin{tabular}{c} $a|\eta|\ll
M^{-1}$\\$p>\frac{1}{3}$
\end{tabular}\big)\\ \hline 0 \big(\begin{tabular}{c} $a|\eta|\ll
M^{-1}$\\$p<\frac{1}{3}$
\end{tabular}\big)
\end{tabular} & \begin{tabular}{c}
$a^{-5}\big(k|\eta|\gg 1\big)$\\ \hline
$a^{-8+\frac{2}{p}}$\big(\begin{tabular}{c} $k|\eta|\ll
1$\\$p>\frac{1}{3}$
\end{tabular}\big)\\ \hline $a^{-2}$ \big(\begin{tabular}{c} $k|\eta|\ll
1$\\$p<\frac{1}{3}$
\end{tabular}\big)
\end{tabular} & \begin{tabular}{c}
$a^{-3}\big(k|\eta|\gg 1\big)$\\ \hline
$a^{-6+\frac{2}{p}}$\big(\begin{tabular}{c} $k|\eta|\ll
1$\\$p>\frac{1}{3}$
\end{tabular}\big)\\ \hline $a^0$ \big(\begin{tabular}{c} $k|\eta|\ll
1$\\$p<\frac{1}{3}$
\end{tabular}\big)
\end{tabular} & \begin{tabular}{c}
$a^{-7+\frac{1}{p}}\big(a|\eta|\gg M^{-1}\big)$\\ \hline
$a^{-\frac{17}{2}+\frac{2}{p}}$\big(\begin{tabular}{c} $a|\eta|\ll
M^{-1}$\\$p>\frac{1}{3}$
\end{tabular}\big)\\ \hline $a^{-\frac{11}{2}+\frac{1}{p}}$ \big(\begin{tabular}{c} $a|\eta|\ll
M^{-1}$\\$p<\frac{1}{3}$
\end{tabular}\big)
\end{tabular}
\\
\hline Sign & \begin{tabular}{c} Negative\\Definite \end{tabular} &
Indefinite & \begin{tabular}{c} Negative\\Definite \end{tabular} &
\begin{tabular}{c} Negative\\Definite \end{tabular} & Indefinite
\\
\hline
\end{tabular}\\
\begin{tabular}{|c|c|c|c|c|}\hline
Terms & $a^{-4}k^2\gamma_0^2+a^{-2}\dot\gamma_0^2$ &
$(-c{\phi_1^{(0)}}^2-d{\phi_2^{(0)}}^2)(a^{-4}k^2\gamma_0^2+a^{-2}\dot\gamma_0^2)$
& $a^{-4}k^2\gamma_0\delta\gamma$ & $a^{-2}\dot\gamma\dot{\delta\gamma}$ \\
\hline Behavior & $a^{-4}$ & \begin{tabular}{c}
$a^{-7}\big(a|\eta|\gg m^{-1}\big)$\\ \hline
$a^{-10+\frac{2}{p}}$\big(\begin{tabular}{c} $a|\eta|\ll
m^{-1}$\\$p>\frac{1}{3}$
\end{tabular}\big)\\ \hline $a^{-4}$ \big(\begin{tabular}{c} $a|\eta|\ll
m^{-1}$\\$p<\frac{1}{3}$
\end{tabular}\big)
\end{tabular} & \begin{tabular}{c}
$a^{-8+\frac{1}{p}}\big(a|\eta|\gg m^{-1}\big)$\\ \hline
$a^{-11+\frac{3}{p}}$\big(\begin{tabular}{c} $a|\eta|\ll
m^{-1}$\\$p>\frac{1}{3}$
\end{tabular}\big)\\ \hline $a^{-4}$ \big(\begin{tabular}{c} $a|\eta|\ll
m^{-1}$\\$p<\frac{1}{3}$
\end{tabular}\big)
\end{tabular} & \begin{tabular}{c}
$a^{-7}\big(a|\eta|\gg m^{-1}\big)$\\ \hline
$a^{-10+\frac{2}{p}}$\big(\begin{tabular}{c} $a|\eta|\ll
m^{-1}$\\$p>\frac{1}{3}$
\end{tabular}\big)\\ \hline $a^{-4}$ \big(\begin{tabular}{c} $a|\eta|\ll
m^{-1}$\\$p<\frac{1}{3}$
\end{tabular}\big)
\end{tabular}
\\
\hline Sign & \begin{tabular}{c} Positive\\Definite \end{tabular} &
\begin{tabular}{c} Indefinite\\(Depending only on $c$ and $d$) \end{tabular} & Indefinite &
Indefinite
\\
\hline
\end{tabular}
\end{widetext}
These terms to not vanish upon spatial averaging.

Note that we have expressed the time dependence in terms of
the dependence on the scale factor $a(t)$. At this stage,
we only need to focus on the exponent of the power-law
scaling. The more negative the power is, the more rapidly the
term grows in a contracting phase (since $a(t)$
is decreasing with time).

As mentioned earlier, the conditions for a bounce to occur in four
space-time dimensional classical Einstein gravity with flat spatial
sections is that the total energy density reaches zero during the
contracting phase. Thus, there needs to be a negative definite
term which starts out small but grows faster than the positive
definite terms due to the regular scalar field and regular radiation.
In the absence of radiation and scalar field inhomogeneities,
it is the contribution to the energy density of the ghost field
$\phi_{2}$ which plays this role.

{F}rom the table we see that there are three kinds of terms: positive
definite, negative definite and indefinite ones. The first set contains
the kinetic and potential terms of the normal scalar as well as the
free energy density of the gauge field, the second set is made up of the
kinetic and potential terms of the ghost scalar, while the third
set contains terms which arise due to the coupling terms between
scalars or between scalars and gauge fields.

Looking first at the terms which are independent of
the coupling term between the fields, we see from the
first line of the ``Behavior'' set of lines in the
third table that, indeed, in the presence of radiation
the energy density in radiation grows faster than that
in the two scalar fields, thus preventing a bounce.
In the presence of coupling between the fields, however,
there are terms
which scale with a larger negative power of $a(t)$.
The signs of some of them, however, depend on the initial phases
for the linear fields $\gamma$, $\phi_{1}^{(1)}$ and
$\phi_{2}^{(1)}$.

Note that the signs of the scalar coupling terms are
determined by the evolution of each field and thus are hard to be
identified in a general analysis. The same is true for
the gauge coupling terms (the last two in the third table).
However, the coupling terms between the scalar fields and the gauge
field (the third to last in the third table) can be made
negative/positive definite easily by setting the signs of
the coefficients $c$ and $d$ to be both positive/negative.

It is reasonable to assume that the contracting phase begins
with the regular scalar field dominating the energy density,
and that the contribution of the Lee-Wick scalar is
much smaller. For single Fourier mode initial conditions
of the radiation field, this can be achieved with the appropriate
choice of the initial phase (see Example 1 in the following
section containing our numerical results). However, for multiple
initial radiation Fourier modes excited any initial phase
difference between the modes will produce a contribution with
the wrong sign and will thus prevent a bounce (see Example 3 in
the following section). In the presence of an infinite set of
modes, the phase correlations required to obtain a bounce
thus appear to have negligible measure in initial condition
space. Thus, even in the presence of coupling between scalar
fields and radiation, the Lee-Wick bounce is unstable.

The bounce, if it exists, will happen at a time which can be
chosen to be $t=0$. Its duration (the time interval lasting
from the time the Hubble radius stops decreasing in the contracting
phase until when it starts expanding in the post-bounce phase) will
be denoted by $\Delta t$. Since the various components of the energy
density scale with different powers of $a(t)$, it is clear that
the duration of the bounce will be shorter or equal to the Hubble
radius $H_{\rm max}^{-1}$ (which gives the time scale on which the ratios of
energy densities in different components change) at the beginning of the
bounce phase. For the background bounce model, we have $H_{\rm max} \sim m$.

There are two kinds of bounces according to the duration of the bounce
phase i) If the period $\Delta t \simeq m^{-1}$, the bounce will go from
the time
\be
t_{B^-} \sim -\frac{(\Delta t)}{2} \sim -\frac{1}{2m}
\ee
to the time
\be
t_{B^+} \sim \frac{(\Delta t)}{2} \sim \frac{1}{2m}
\ee
with a low speed. We call this a ``slow bounce''. In this case,
the universe will enter the bounce period at the critical time
$t_c\sim m$, and only the $a|\eta|\gg m^{-1}$ approximate solutions
of the previous tables will be applicable and not ones for the interval
$a|\eta|\ll m^{-1}$. ii) If the period $\Delta t \ll m^{-1}$,
the bounce will happen in a very short time with very fast speed. This
can be called the ``fast bounce''. In this case, the universe evolves
from the far past ($-t_i$ with $|t_i|\gg 1$) to $t=0$, passing through
the point $t_c \sim m$, then entering into the region $|t|\lesssim m$
before finally reaching the bounce point. In this case, both of the
two approximate solutions of the field evolution will be applied.

Let us now consider the necessary conditions for a bounce (as we
have indicated above and will see from the numerical analysis, these
conditions are not sufficient - in addition to the conditions which
follow, appropriate correlations in the initial phases are required).
We start in the region of time $a|\eta|\gg m^{-1}$. We study
the conditions required to have the terms that might give a bounce
grow relative to the other terms during this phase. If the
conditions are not satisfied, or the bounce does not happen even
if the conditions are satisfied, then a bounce may still occure
in the $a|\eta|\ll m^{-1}$ region. The conditions for
the terms in the energy which could compensate the positive radiation
contribution to become dominant are then studied. If these
conditions are not satisfied, either, then a bounce is impossible.

A necessary condition for a bounce to be possible requires the
growth rate of one of the indefinite sign terms in the third table
above exceed all that of all of the positive definite terms. In the
$a|\eta|\gg m^{-1}$ region, this requires $8-1/p>5$, which
equivalently constrains the equation of state parameter $w$ to be in
the range $w<1$. If this condition is satisfied in this region, then
a slow bounce may happen depending on the choice of the initial
phases.

If the condition is not satisfied in the $a|\eta|\gg m^{-1}$ region,
the universe may evolve into the $a|\eta|\ll m^{-1}$ region, in
which the evolution of the fields are different, and new constraints
on $p$ and $w$ will arise if a bounce is to be possible. Following
the above logic, we find that the conditions under which a bounce
might happen are much looser, namely $w > - 7/6$.

To summarize this section: we have identified necessary conditions
for a bounce to occur. Whether one actually does occur even if
the conditions are satisfied depends on the initial phases of
the fields. This must be studied numerically.
In the following section we will give one example of specially chosen
phases for which a bounce is possible. However, when we
look at a more general choice of phases, the bounce will not
occur.

\section{Numerical Results}

In order to support the analysis in the last section, we
performed numerical calculations. Such numerical work is
necessary because our analytical analysis is only approximate.
In particular, we worked in perturbation theory up to order
second order in $\e$. In addition, even in cases where
our analytical analysis would indicate the possibility of
a bounce, the perturbative analysis will break down near
the bounce point, and there is no assurance that the trends
seen in the perturbative analysis will persist.

We have numerically solved the full nonlinear equations of
motion for the matter fields in the presence of a homogeneous
expanding background cosmology. The homogeneous cosmology
is obtained numerically by solving the first Friedmann equation
\be
H^2 = \frac{8 \pi G}{3} \rho \, ,
\ee
where $G$ is Newton's gravitational constant (related to the Planck
mass used earlier), and $\rho$ is the total energy density, averaged
over space.

\begin{figure}[htbp]
\includegraphics[scale=0.25]{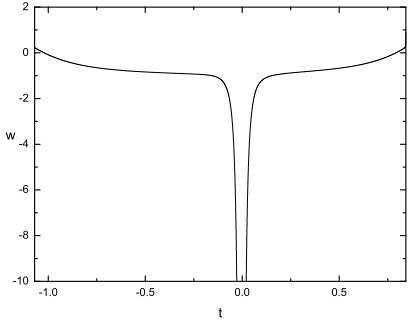}
\caption{The evolution of the equation of state $w$ w.r.t. cosmic time
$t$ (horizontal axis), in the first simulation, a simulation with only
a single Fourier mode excited and phases chosen as indicated below.
We see that $w$ drops to $-\infty$, indicating
that there is a bounce. The background fields are plotted in
dimensionless units by normalizing by the mass $m_{rec}=10^{-6}
m_{Pl}$ while the time axis is displayed in units of $m^{-1}_{rec}$.
The mass parameters m and M were chosen to be $m=5m_{rec}$ and
$M=10m_{rec}$. The initial conditions were chosen to be
$\gamma_i\simeq-1.85\times10^5m_{rec}$,
$\dot\gamma_i\simeq7.35\times10^6m^2_{rec}$,
$\phi_{1i}\simeq1.015\times10^5m_{rec}$,
$\dot\phi_{1i}\simeq6.39\times10^5m^2_{rec}$,
$\phi_{2i}\simeq2.54\times10^2m_{rec}$,
$\dot\phi_{2i}\simeq-4.96\times10^3m^2_{rec}$. The coefficients $c$
and $d$ are chosen to be $c=10^{-10}M_{rec}^2$ and
$d=10^{-10}M_{rec}^2$. The wavenumber is $k\simeq 0.01h Mpc^{-1}$.
}\label{1}
\end{figure}

\begin{figure}[htbp]
\includegraphics[scale=0.25]{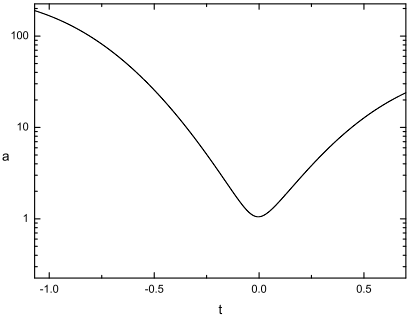}
\caption{The scale factor of the universe in the same simulation
that leads to the evolution of the equation of state shown in
Fig. \ref{1}. From the plot we see that the
bounce happens at $t=0$.}\label{2}
\end{figure}

\begin{figure}[htbp]
\includegraphics[scale=0.25]{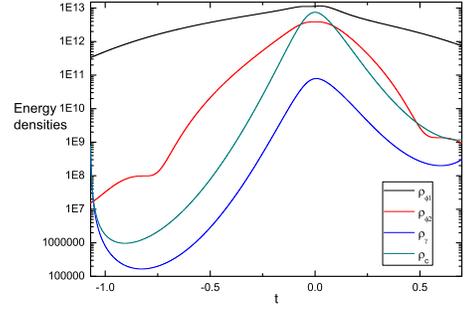}
\caption{(Color online.) Energy densities of $\phi_1$, $\phi_2$ and
$\gamma$ in the system with parameters chosen as in Fig. \ref{1}.
The curves from up to down are: $\rho_{\phi_1}$ (black),
$\rho_{\phi_2}$ (red), $\rho_c$ (dark cyan) and $\rho_\gamma$
(blue), respectively. The variables are also normalized with the
mass scale $m_{rec}=10^{-6} m_{Pl}$. } \label{3}
\end{figure}

\begin{figure}[htbp]
\includegraphics[scale=0.25]{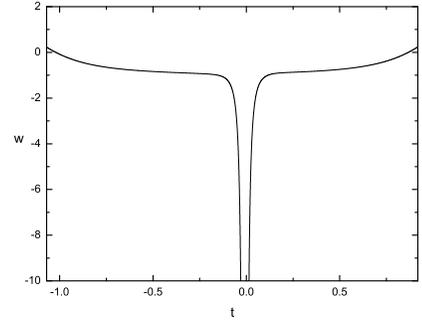}
\caption{The evolution of the equation of state $w$ as a function of cosmic time
$t$ (horizontal axis). The behavior that $w$ drops to $-\infty$
indicating that a bounce takes place. The background fields are plotted in
dimensionless units by normalizing by the mass $m_{rec}=10^{-6}
m_{Pl}$ while the time axis is displayed in units of $m^{-1}_{rec}$.
The mass parameters $m$ and $M$ were chosen to be $m=5m_{rec}$ and
$M=10m_{rec}$. The initial conditions were chosen to be
$\gamma_i\simeq-1.85\times10^5m_{rec}$,
$\dot\gamma_i\simeq7.35\times10^6m^2_{rec}$,
$\phi_{1i}\simeq1.015\times10^5m_{rec}$,
$\dot\phi_{1i}\simeq6.39\times10^5m^2_{rec}$,
$\phi_{2i}\simeq2.54\times10^2m_{rec}$,
$\dot\phi_{2i}\simeq-4.96\times10^3m^2_{rec}$. The coefficients $c$
and $d$ are chosen to be $c=10^{-10}M_{rec}^2$ and
$d=-10^{-10}M_{rec}^2$. The wavenumber $k\simeq 0.01h Mpc^{-1}$.
}\label{4}
\end{figure}

\begin{figure}[htbp]
\includegraphics[scale=0.25]{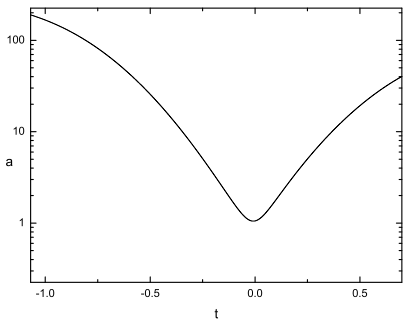}
\caption{The scale factor of the universe driven by the system with
parameters chosen as in Fig. \ref{4}. From the plot we see that the
bounce happens at $t=0$.}\label{5}
\end{figure}

\begin{figure}[htbp]
\includegraphics[scale=0.25]{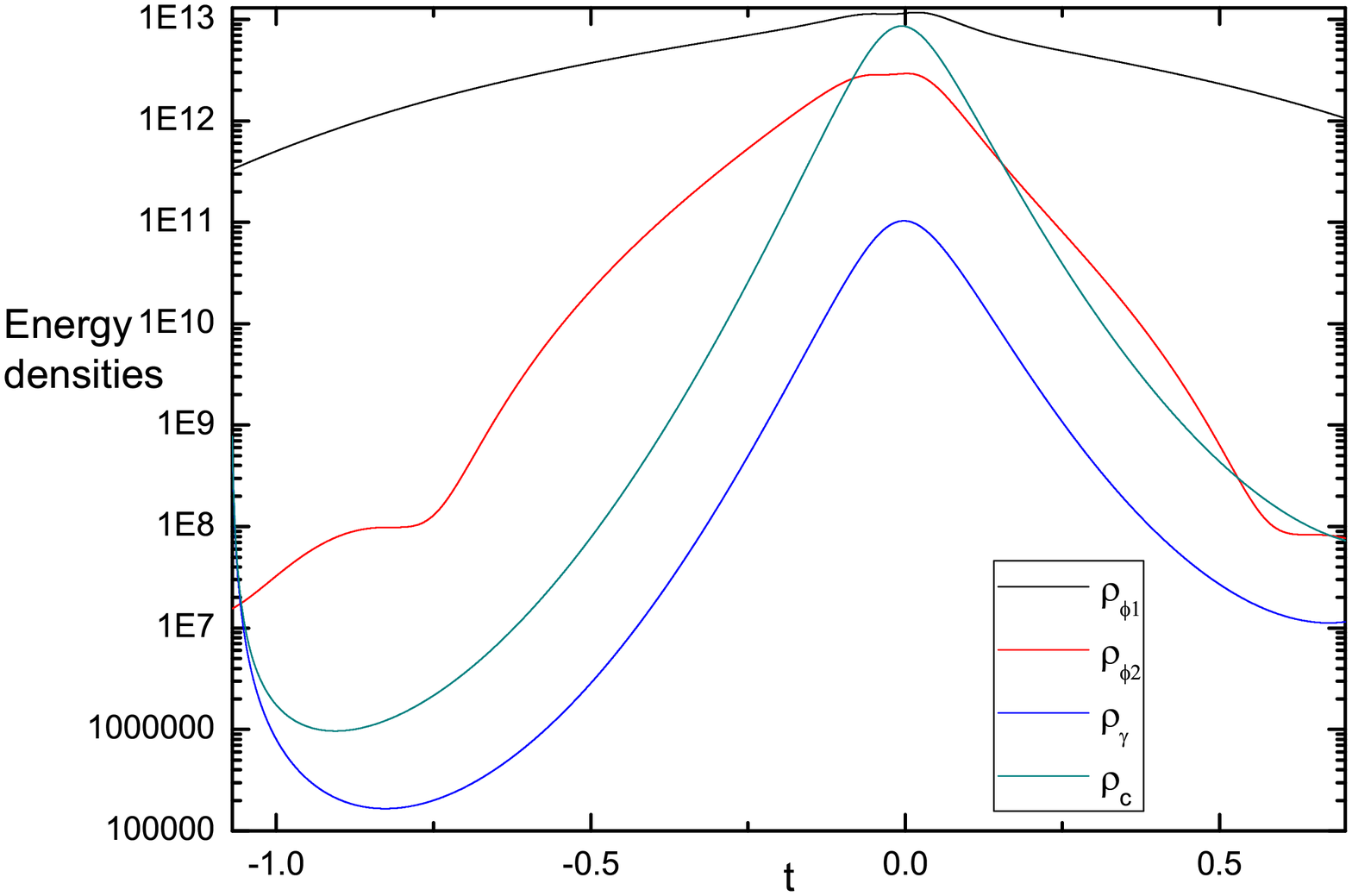}
\caption{(Color online.) Energy densities of $\phi_1$, $\phi_2$ and
$\gamma$ in the system with parameters chosen as in Fig. \ref{4}.
The curves from up to down are: $\rho_{\phi_1}$ (black),
$\rho_{\phi_2}$ (red), $\rho_c$ (dark cyan) and $\rho_\gamma$
(blue), respectively. The variables are also normalized with the
mass scale $m_{rec}=10^{-6} m_{Pl}$. }\label{6}
\end{figure}

Figures \ref{1}-\ref{6} are two groups of numerical results with
different parameters. In both cases we choose the initial energy
density of the gauge fields to be larger than that of the Lee-Wick
scalar, but less than that of the normal scalar. These initial
conditions correspond to the situation we are interested in, namely
starting in a matter-dominated contracting phase in the presence of
some radiation which is sub-dominant. Fig. \ref{1}, \ref{2} and
\ref{3} show an example with parameters $c>0$ and $d>0$. We choose
initial conditions in which a single Fourier mode fluctuation is
excited, and in which the phases are chosen as indicated in the
figure caption. For these initial phases, we obtain a bounce. In
Fig. \ref{1}, we see that the equation of state $w$ begins with a
value slightly larger than $0$, and then evolves to some nearly
fixed value. For the case of our initial condition choice, it
appears to be $w  \simeq -0.6$, in the region where the bounce is
allowed to happen.
At the bounce point,
the equation of state will drop to $-\infty$, while after the
bounce, the equation of state will rise again to $w \simeq 0.6$.
Fig. \ref{2} is the plot of the scale factor in this case
which shows explicitly the occurrence of the bounce.
Fig. \ref{3} gives a comparison of the energy densities of some
components during the process. Initially, we set the energy density
of the gauge field $\gamma$ to be between the normal scalar and
Lee-Wick scalar. When the evolution of the universe enters into a
region with nearly constant $w$, the gauge-coupling component of
energy density $\rho_c$ will grow very fast. It is negative and thus
enables the negative part of the energy density to catch up
with the positive one, thus allowing the bounce to happen. For the
inhomogeneous fluctuation, we choose the wavenumber to be
$k\simeq 0.01h Mpc^{-1}$ which corresponds to a scale which
is observable by CMB and LSS experiments.

Figs. \ref{4}, \ref{5} and \ref{6} give the corresponding results
in the case when we choose $c>0$ while $d<0$ (with all initial
conditions identical). This case seems dangerous because the
contribution of the Lee-Wick scalar to the fluctuation terms
could lead to an instability. However, as we have
mentioned before, since the effects of the Lee-Wick scalar are less than
that of the normal scalar, it is still possible for the bounce to
happen. Fig. \ref{4} shows the equation of state of the system. We can
see that the evolution of $w$ is about the same as that in Fig.
\ref{1}, since the change of the sign of $d$ does not alter the
result too much. Fig. \ref{5} is the behavior of scale factor in this
case while Fig. \ref{6} gives the comparison of the energy densities of
all components.

A change in the phase of the initial radiation field velocity will not change
the results (if we keep the other initial conditions fixed). On the other
hand, if we flip the sign of the initial velocity of one of the two scalar
fields, then the sign of the dominant contribution to the energy density
as we approach the bounce will flip and this will prevent a bounce. If we
use initial conditions containing two excited Fourier modes, then we obtain
a bounce only if the signs of the initial field velocities are both chosen
as in the first run whose results are shown here. Different phases for the
scalar field velocities of the two modes destroys the possibility of
obtaining a bounce.

\begin{figure}[htbp]
\includegraphics[scale=0.35]{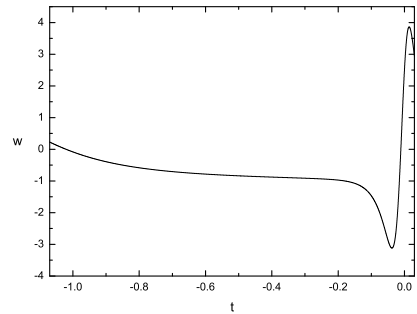}
\caption{The evolution of the equation of state $w$ as a function of
cosmic time $t$ (horizontal axis). The background fields are plotted
in dimensionless units by normalizing by the mass $m_{rec}=10^{-6}
m_{Pl}$ while the time axis is displayed in units of $m^{-1}_{rec}$.
The mass parameters $m$ and $M$ were chosen to be $m=5m_{rec}$ and
$M=10m_{rec}$. This plot is the evolution of the system with two
Fourier modes combined together. The one is of which the wavenumber
$k\simeq 0.01h Mpc^{-1}$ with initial conditions
$\gamma_i\simeq-1.85\times10^5m_{rec}$,
$\dot\gamma_i\simeq7.35\times10^6m^2_{rec}$,
$\phi_{1i}\simeq1.015\times10^5m_{rec}$,
$\dot\phi_{1i}\simeq6.39\times10^5m^2_{rec}$,
$\phi_{2i}\simeq2.54\times10^2m_{rec}$,
$\dot\phi_{2i}\simeq-4.96\times10^3m^2_{rec}$, which if taken alone
will give the bounce as has been shown in the previous example. The
other is of which the wavenumber $k\simeq 0.04h Mpc^{-1}$ with
initial conditions of the same initial values of the fields but the
opposite signs of the scalar field velocity. From this plot we can
see that the combination of the two Fourier mode will (generally)
cause $w$ blow up, thus preventing the bounce. This means that the
bounce requires special fine-tuning of the initial phases for each
Fourier mode. The coefficients $c$ and $d$ are chosen to be
$c=10^{-10}M_{rec}^2$ and $d=-10^{-10}M_{rec}^2$. }\label{7}
\end{figure}

\begin{figure}[htbp]
\includegraphics[scale=0.35]{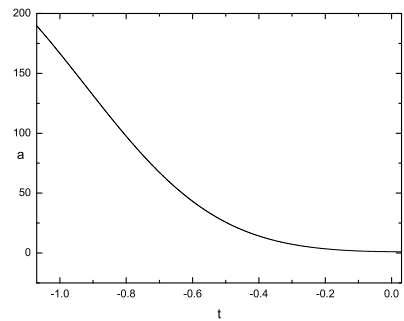}
\caption{The Hubble constant of the universe driven by the system
with parameters chosen as in Fig. \ref{7}. From the plot we see that
there is a singularity at $t=0$.}\label{8}
\end{figure}

\begin{figure}[htbp]
\includegraphics[scale=0.35]{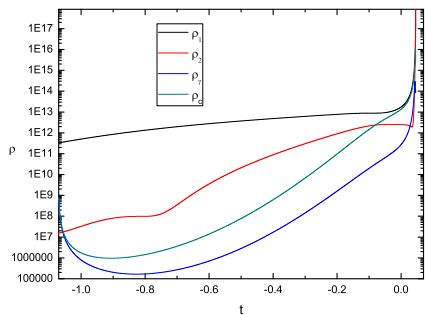}
\caption{(Color online.) Energy densities of $\phi_1$, $\phi_2$ and
$\gamma$ in the system with parameters chosen as in Fig. \ref{7}.
The curves from up to down are: $\rho_{\phi_1}$ (black),
$\rho_{\phi_2}$ (red), $\rho_c$ (dark cyan) and $\rho_\gamma$
(blue), respectively. The variables are also normalized with the
mass scale $m_{rec}=10^{-6} m_{Pl}$. }\label{9}
\end{figure}

Figures \ref{7}, \ref{8} and \ref{9} show the results for the
equation of state parameter $w$, the Hubble parameter $H$ and the
contribution of the various components to the total $\rho$ in
the case of a simulation in which two Fourier modes are excited,
with velocities of both scalar fields having opposite signs from
those in the previous example. As is obvious, a Big Crunch
singularity occurs.


\section{Conclusions and Discussion}

In this paper we analyzed in detail the possibility of obtaining a
cosmological bounce in a model which corresponds to the scalar field
sector of the Lee-Wick theory coupled to relativistic radiation. It
is known that the scalar field sector of the Lee-Wick theory in the
absence of other fields can yield a cosmological bounce
\cite{Cai:2008qw}. In fact, the universe will scale as
non-relativistic matter with $<w>\simeq 0$ both before and after the
bounce. Thus, this model is a possible realization of the ``matter
bounce" scenario. However, this background is unstable to the
introduction of radiation since in the contracting phase the growth
of energy density in radiation will exceed that of matter and will
lead to a Big Crunch singularity As has been shown in previous work
\cite{Karouby:2010wt}, the introduction of a Lee-Wick partner to
radiation does not prevent this instability. In this paper, we
introduced an interaction between the radiation field and the scalar
fields. The interaction could help drain energy from the radiation
field to the Lee-Wick scalar, and thus could prevent the radiation
from growing too fast to destroy the bounce.

We analyzed the equations describing the evolution of the three matter
fields (regular scalar field, its Lee-Wick partner and the radiation field)
on a cosmological background both analytically and numerically.
Our analytical analysis was perturbative and made use of the second
order using Born approximation. The expansion parameter is set by the
initial amplitude of the gauge field. We solved the
equations of motion for each field at each order, and obtained their
approximations in different cases. We compared their contributions
to the total energy density, and derived necessary conditions for a bounce to
happen. To support our analysis, we also performed numerical
calculations.

Specifically, we investigated initial conditions in which one or two
Fourier modes of the radiation field and the scalar field fluctuations
are excited. We found special initial conditions which indeed lead
to a non-singular bounce. Changing the sign of the initial scalar
field velocity will destroy the bounce solution. In the presence of
two Fourier modes, we found that a bounce requires identical initial
phases for the two modes. For general initial conditions, we
conjecture that the measure of such initial conditions which lead
to a bounce is very small. We thus find that the addition of coupling
terms between the scalar fields and radiation cannot save the
Lee-Wick bounce background from the instability problem with
respect to the addition of radiation (nor, for that matter, with
respect to scalar field fluctuations). The instability problem with
respect to anisotropic stress will be even worse.

\begin{acknowledgments}

The work at McGill is supported in part by an NSERC Discovery Grant
and by funds from the Canada Research Chair program. RB is recipient
of a Killam Research Fellowship. The work at CYCU is funded in parts
by the National Science Council of R.O.C. under Grant No.
NSC99-2112-M-033-005-MY3 and No. NSC99-2811-M-033-008 and by the
National Center for Theoretical Sciences.

\end{acknowledgments}

\appendix
\section{Green's function determination of $u_i ^{(2)}$}

The solution for the second order scalar field correction
$u_{i}^{(2)}$ can be determined using the Green function
method. The general solution of (\ref{eqn2new}) is the sum of
the solution $u_0(\eta)$ of the homogeneous equation which solves
the same initial conditions as $u$ and the particular solution
$\delta u(\eta)$ wich vanishes at time $\eta_I$.  The particular
solution is given by
\ba \label{GF} \delta u(\eta) &=&
u_a(\eta)\int_{\ei}^{\eta}d\eta'\epsilon(\eta')u_b(\eta')s(\eta')\nonumber\\
&&-u_b(\eta)\int_{\ei}^{\eta}d\eta'\epsilon(\eta')u_a(\eta')s(\eta')~,
\ea 
where $u_1$ and $u_2$ are two independent solutions of the
homogeneous equation, $\epsilon(\eta)$ is the Wronskian:
\be
\epsilon(\eta) \, = \, \bigl( u_a^{'} u_b - u_b^{'} u_a \bigr)^{-1}
\, ,
\ee
and $s(\eta)$ is the source inhomogeneity.

Recall from the main text that the second order field
correction terms satisfy the equations:
\ba\label{eqn3new} \left\{ \begin{array}{c}

{u_1^{(2)}}''+a^2m^2u_1^{(2)} = c\frac{(k^2\gamma^2-\gamma'^2)}{a^2}u_1^{(0)}~,\\
\\
{u_2^{(2)}}''+a^2M^2u_2^{(2)} = -d\frac{(k^2\gamma^2-\gamma'^2)}{a^2}u_2^{(0)}~,\\
\end{array}\right.
\ea

We will demonstrate the analysis for the case of $u_1^{(2)}$.
Let us consider evolution for a short interval of time starting
at some initial time $\eta_I$. Then, we can neglect the
expansion of the universe in the equation of motion
and take $a(\eta)=a(\eta_I)$. We are then interested in
how the result scales in $\eta_I$. Using this
trick, the solutions of the homogeneous equation can
be taken to be
\ba
u_a(\eta) &=& \cos(\om_m \eta) \nonumber \\
u_b(\eta) &=& \sin(\om_m \eta) \ea 
and the Wronskian is
\be
\epsilon(\eta) = -\frac{ 1}{\om_m} \,\,\, {\rm where}
\,\,\, \om_m = \sqrt {a^2 m^2} \, .
\ee

Using the result for the background $\gamma$ from
the main text, the source term becomes
\ba
s_\ga (\eta)\,&=& c\frac{(k^2\gamma^2-\gamma'^2)}{a^2}u_1^{(0)}\nonumber\\
              & \sim & a^{-2}k^2
              |\eta|^\frac{p}{2(p-1)}\times \\
              &&\cos(2k|\eta|+2\te_\gamma)\cos((1-p)am|\eta|+\te_1^{(0)})~,
\nonumber
\ea
since $\gamma_0 \sim \cos(k|\eta|+\te_\gamma)$.

Combining these results we obtain 
\ba \label{GF3} u_1^{(2)} &\sim&
-\cos(\omega_{m}\eta)\int_{\ei}^{\eta} \frac{d\eta k^2}
{a^3(t)m}|\eta|^\frac{p}{2(p-1)}\times\nonumber\\
&&\sin(\omega_{m}\eta)\cos(2k|\eta|+2\te_\gamma)\cos((1-p)am|\eta|+\te_1^{(0)}) \nonumber \\
 & & + \sin(\omega_{m}\eta)\int_{\ei}^{\eta} \frac{d\eta k^2}
{a^3(t)m}|\eta|^\frac{p}{2(p-1)}\times \\
&&\cos(\omega_{m}\eta)\cos(2k|\eta|+2\te_\gamma)\sin((1-p)am|\eta|+\te_1^{(0)})~, \nonumber \\
u_2^{(2)} &\sim& -\cos(\omega_{M}\eta)\int_{\ei}^{\eta} \frac{d\eta
k^2}
{a^3(t)M}|\eta|^\frac{p}{2(p-1)}\times\nonumber\\
&&\sin(\omega_{M}\eta)\cos(2k|\eta|+2\te_\gamma)\cos((1-p)aM|\eta|+\te_1^{(0)}) \nonumber \\
 & & + \sin(\omega_{M}\eta)\int_{\ei}^{\eta} \frac{d\eta k^2}
{a^3(t)M}|\eta|^\frac{p}{2(p-1)}\times \\
&&\cos(\omega_{M}\eta)\cos(2k|\eta|+2\te_\gamma)\sin((1-p)aM|\eta|+\te_1^{(0)})~.
\nonumber
\ea
Note that if we only care about their scalings with respect to
conformal time $\eta$ or scale factor $a(t)$, the above solutions
can be reduced to:
\be
u_{1,2}^{(2)}\propto
|\eta|^{\frac{9p-2}{2(p-1)}}\propto
a^{-\frac{9}{2}+\frac{1}{p}}~,
\ee
and for the case of a matter
dominated era where $p=2/3$, it is straightforward to show that
\be
u_{1,2}^{(2)}\propto |\eta|^{-6}\propto a^{-3}~.
\ee




\begin{thebibliography}{99}

\bibitem{Cai:2008qw}
  Y.~F.~Cai, T.~t.~Qiu, R.~Brandenberger and X.~m.~Zhang,
  ``A Nonsingular Cosmology with a Scale-Invariant Spectrum of Cosmological
  Perturbations from Lee-Wick Theory,''
  Phys.\ Rev.\  D {\bf 80} (2009) 023511
  [arXiv:0810.4677 [hep-th]].

 \bibitem{Hawking:1973uf}
  S. W. Hawking, and G. F. R. Ellis, {\it The large scale structure
  of space-time}, Cambridge University Press (1973); A.~Borde and
  A.~Vilenkin, Phys.\ Rev.\ Lett.\  {\bf 72}, 3305 (1994).

\bibitem{Borde:1993xh}
  A.~Borde and A.~Vilenkin,
  ``Eternal Inflation And The Initial Singularity,''
  Phys.\ Rev.\ Lett.\  {\bf 72}, 3305 (1994)
  [arXiv:gr-qc/9312022].

\bibitem{Mukhanov:1991zn}
V.~F.~Mukhanov and R.~H.~Brandenberger,
  ``A Nonsingular universe,''
  Phys.\ Rev.\ Lett.\  {\bf 68}, 1969 (1992);\\
  R.~H.~Brandenberger, V.~F.~Mukhanov and A.~Sornborger,
  ``A Cosmological theory without singularities,''
  Phys.\ Rev.\  D {\bf 48}, 1629 (1993)
  [arXiv:gr-qc/9303001].

 \bibitem{Novello}
   M.~Novello and S.~E.~P.~Bergliaffa,
  ``Bouncing Cosmologies,''
  Phys.\ Rept.\  {\bf 463}, 127 (2008)
  [arXiv:0802.1634 [astro-ph]].

\bibitem{Wands}
 D.~Wands,
  ``Duality invariance of cosmological perturbation spectra,''
  Phys.\ Rev.\  D {\bf 60}, 023507 (1999)
  [arXiv:gr-qc/9809062].

\bibitem{Finelli}
 F.~Finelli and R.~Brandenberger,
  ``On the generation of a scale-invariant spectrum of adiabatic  fluctuations
  in cosmological models with a contracting phase,''
  Phys.\ Rev.\  D {\bf 65}, 103522 (2002)
  [arXiv:hep-th/0112249].

\bibitem{models}
P.~Peter and N.~Pinto-Neto,
  ``Primordial perturbations in a non singular bouncing universe model,''
  Phys.\ Rev.\  D {\bf 66}, 063509 (2002)
  [arXiv:hep-th/0203013];\\
 F.~Finelli,
  ``Study of a class of four dimensional nonsingular cosmological bounces,''
  JCAP {\bf 0310}, 011 (2003)
  [arXiv:hep-th/0307068];\\
 L.~E.~Allen and D.~Wands,
  ``Cosmological perturbations through a simple bounce,''
  Phys.\ Rev.\  D {\bf 70}, 063515 (2004)
  [arXiv:astro-ph/0404441];\\
R.~Brandenberger, H.~Firouzjahi and O.~Saremi,
  ``Cosmological Perturbations on a Bouncing Brane,''
  JCAP {\bf 0711}, 028 (2007)
  [arXiv:0707.4181 [hep-th]];\\
S.~Alexander, T.~Biswas and R.~H.~Brandenberger,
  ``On the Transfer of Adiabatic Fluctuations through a Nonsingular
  Cosmological Bounce,''
  arXiv:0707.4679 [hep-th];\\
 F.~Finelli, P.~Peter and N.~Pinto-Neto,
  ``Spectra of primordial fluctuations in two-perfect-fluid regular bounces,''
  Phys.\ Rev.\  D {\bf 77}, 103508 (2008)
  [arXiv:0709.3074 [gr-qc]];\\
A.~Cardoso and D.~Wands,
  ``Generalised perturbation equations in bouncing cosmologies,''
  Phys.\ Rev.\  D {\bf 77}, 123538 (2008)
  [arXiv:0801.1667 [hep-th]];\\
Y.~F.~Cai, T.~Qiu, R.~Brandenberger, Y.~S.~Piao and X.~Zhang,
  ``On Perturbations of Quintom Bounce,''
  JCAP {\bf 0803}, 013 (2008)
  [arXiv:0711.2187 [hep-th]];\\
X.~Gao, Y.~Wang, W.~Xue and R.~Brandenberger,
  ``Fluctuations in a Ho\v{r}ava-Lifshitz Bouncing Cosmology,''
  JCAP {\bf 1002}, 020 (2010)
  [arXiv:0911.3196 [hep-th]];\\
  T.~Qiu and K.~C.~Yang,
  ``Perturbations in Matter Bounce with Non-minimal Coupling,''
  JCAP {\bf 1011}, 012 (2010)
  [arXiv:1007.2571 [astro-ph.CO]];\\
C.~Lin, R.~H.~Brandenberger and L.~P.~Levasseur,
  ``A Matter Bounce By Means of Ghost Condensation,''
  arXiv:1007.2654 [hep-th].

\bibitem{RHBrev0}
R.~H.~Brandenberger,
  ``Inflationary cosmology: Progress and problems,''
  arXiv:hep-ph/9910410.

\bibitem{Martin:2000xs}
J.~Martin and R.~H.~Brandenberger,
``The TransPlanckian problem of inflationary cosmology,''
Phys.~Rev.~D~{\bf 63}, 123501 (2001), [arXiv:hep-th/0005209];\\
R.~H.~Brandenberger and J.~Martin,
  ``The robustness of inflation to changes in super-Planck-scale physics,''
  Mod.\ Phys.\ Lett.\  A {\bf 16}, 999 (2001)
  [arXiv:astro-ph/0005432].

\bibitem{Cai:2007qw}
  Y.~F.~Cai, T.~Qiu, Y.~S.~Piao, M.~Li and X.~Zhang,
  ``Bouncing Universe with Quintom Matter,''
  JHEP {\bf 0710}, 071 (2007)
  [arXiv:0704.1090 [gr-qc]].

\bibitem{LW}
T.~D.~Lee and G.~C.~Wick,
  ``Negative Metric and the Unitarity of the S Matrix,''
  Nucl.\ Phys.\  B {\bf 9}, 209 (1969);\\
T.~D.~Lee and G.~C.~Wick,
  ``Finite Theory of Quantum Electrodynamics,''
  Phys.\ Rev.\  D {\bf 2}, 1033 (1970);\\
B.~Grinstein, D.~O'Connell and M.~B.~Wise,
  ``The Lee-Wick standard model,''
  Phys.\ Rev.\  D {\bf 77}, 025012 (2008)
  [arXiv:0704.1845 [hep-ph]].

\bibitem{Ekp}
J.~Khoury, B.~A.~Ovrut, P.~J.~Steinhardt and N.~Turok,
  ``The ekpyrotic universe: Colliding branes and the origin of the hot big
  bang,''
  Phys.\ Rev.\  D {\bf 64}, 123522 (2001)
  [arXiv:hep-th/0103239].

\bibitem{Jeon}
J.~M.~Cline, S.~Jeon and G.~D.~Moore,
  ``The phantom menaced: Constraints on low-energy effective ghosts,''
  Phys.\ Rev.\  D {\bf 70}, 043543 (2004)
  [arXiv:hep-ph/0311312].

\bibitem{Karouby:2010wt}
  J.~Karouby and R.~Brandenberger,
  ``A Radiation Bounce from the Lee-Wick Construction?,''
  Phys.\ Rev.\  D {\bf 82}, 063532 (2010)
  [arXiv:1004.4947 [hep-th]].


\end{thebibliography}
\end{document}